\documentclass[9pt,
aps,prl,
 superscriptaddress,
]{revtex4-1}
\usepackage{dcolumn}
\usepackage{bm}
\usepackage{xr}
\usepackage%
{graphicx}
\usepackage[normalem]{ulem}
\usepackage{xfrac}
\usepackage[nice]{nicefrac}
\usepackage{extdash}
\usepackage{epsfig}
\usepackage{isomath}
\usepackage{textcomp}
\usepackage{bm}
\usepackage[english]{babel}
\usepackage{todonotes}
\usepackage{color,soul}
\usepackage{caption}
\usepackage{verbatim}
\usepackage{times}

\DeclareFontFamily{U}{euc}{}
\DeclareFontShape{U}{euc}{m}{n}{<-6>eurm5<6-8>eurm7<8->eurm10}{}%
\DeclareSymbolFont{AMSc}{U}{euc}{m}{n} 
\DeclareMathSymbol{\umu}{\mathord}{AMSc}{"16} 
\renewcommand{\vec}[1]{\boldsymbol{#1}}
\newcommand{\ensuretext}[1]{\ensuremath{\text{#1}}}

\newcommand{\unit}[1]{\ensuretext{\textrm{\,}}\ensuremath{\mathrm{#1}}}
\newcommand{\eV}{\mathrm{eV}}

\newcommand{\keV}{\mathrm{k}\eV}

\newcommand{\Mum}{\ensuremath{\umu}\ensuremath{\mathrm{m}}}
\newcommand{\mum}{\textrm{\,\ensuremath{\mathrm{\Mum}}}}
\newcommand{\eqref}[1]{(\ref{#1})}

\mathchardef\ordinarycolon\mathcode`\:
\mathcode`\:=\string"8000
\begingroup \catcode`\:=\active
  \gdef:{\mathrel{\mathop\ordinarycolon}}
\endgroup

\begin{document}

\title{Probing ultrafast laser plasma processes inside solids with resonant small-angle X-ray scattering} 
\author{Lennart Gaus}
\affiliation{Helmholtz-Zentrum Dresden-Rossendorf, Bautzner Landstra\ss e 400, 01328, Dresden, Germany}
\affiliation{Technische Universität Dresden, 01069 Dresden, Germany}
\author{Lothar Bischoff}
\affiliation{Helmholtz-Zentrum Dresden-Rossendorf, Bautzner Landstra\ss e 400, 01328, Dresden, Germany}
\author{Michael Bussmann}
\affiliation{Helmholtz-Zentrum Dresden-Rossendorf, Bautzner Landstra\ss e 400, 01328, Dresden, Germany}
\affiliation{Center for Advanced Systems Understanding (CASUS), Görlitz, Germany}
\author{Eric Cunningham}
\affiliation{SLAC National Accelerator Laboratory, 2575 Sand Hill Rd, Menlo Park, CA 94025, USA}
\author{Chandra B.  Curry}
\affiliation{SLAC National Accelerator Laboratory, 2575 Sand Hill Rd, Menlo Park, CA 94025, USA}
\affiliation{University of Alberta, 116 St. and 85 Ave. Edmonton, AB T6G 2R3, Canada}
\author{Eric Galtier}
\affiliation{SLAC National Accelerator Laboratory, 2575 Sand Hill Rd, Menlo Park, CA 94025, USA}
\author{Maxence Gauthier}
\affiliation{SLAC National Accelerator Laboratory, 2575 Sand Hill Rd, Menlo Park, CA 94025, USA}
\author{Alejandro Laso Garc\'ia}
\affiliation{Helmholtz-Zentrum Dresden-Rossendorf, Bautzner Landstra\ss e 400, 01328, Dresden, Germany}
\author{Marco Garten}
\affiliation{Helmholtz-Zentrum Dresden-Rossendorf, Bautzner Landstra\ss e 400, 01328, Dresden, Germany}
\affiliation{Technische Universität Dresden, 01069 Dresden, Germany}
\author{Siegfried Glenzer}
\affiliation{SLAC National Accelerator Laboratory, 2575 Sand Hill Rd, Menlo Park, CA 94025, USA}
\author{J\"org Grenzer}
\affiliation{Helmholtz-Zentrum Dresden-Rossendorf, Bautzner Landstra\ss e 400, 01328, Dresden, Germany}
\author{Christian Gutt}
\affiliation{Universit\"at Siegen, Department Physik, Walter-Flex-Stra\ss e 3, 57072 Siegen, Germany}
\author{Nicholas J. Hartley}
\affiliation{Helmholtz-Zentrum Dresden-Rossendorf, Bautzner Landstra\ss e 400, 01328, Dresden, Germany}
\affiliation{SLAC National Accelerator Laboratory, 2575 Sand Hill Rd, Menlo Park, CA 94025, USA}
\author{Lingen Huang}
\affiliation{Helmholtz-Zentrum Dresden-Rossendorf, Bautzner Landstra\ss e 400, 01328, Dresden, Germany}
\author{Uwe H\"ubner}
\affiliation{Leibniz Institute of Photonic Technology, Albert-Einstein-Stra\ss e 9, 07745 Jena, Germany}
\author{Dominik Kraus}
\affiliation{Helmholtz-Zentrum Dresden-Rossendorf, Bautzner Landstra\ss e 400, 01328, Dresden, Germany}
\author{Hae Ja Lee}
\affiliation{SLAC National Accelerator Laboratory, 2575 Sand Hill Rd, Menlo Park, CA 94025, USA}
\author{Emma E. McBride}
\affiliation{SLAC National Accelerator Laboratory, 2575 Sand Hill Rd, Menlo Park, CA 94025, USA}
\affiliation{European XFEL, Holzkoppel 4, 22869 Schenefeld, Germany}
\author{Josefine Metzkes-Ng}
\affiliation{Helmholtz-Zentrum Dresden-Rossendorf, Bautzner Landstra\ss e 400, 01328, Dresden, Germany}
\author{Bob Nagler}
\affiliation{SLAC National Accelerator Laboratory, 2575 Sand Hill Rd, Menlo Park, CA 94025, USA}
\author{Motoaki Nakatsutsumi}
\affiliation{European XFEL, Holzkoppel 4, 22869 Schenefeld, Germany}
\author{Jan Nikl}
\affiliation{ELI Beamlines Centre, Institute of Physics, Czech Academy of Sciences, Za Radnicí 835, Dolní Břežany 25241, Czech Republic}
\affiliation{Faculty of Nuclear Sciences and Physical Engineering, Czech Technical University in Prague, Břehová 7, Prague 11519, Czech Republic}
\affiliation{Institute of Plasma Physics, Czech Academy of Sciences, Za Slovankou 1782/3, Prague 18200, Czech Republic}
\author{Masato Ota}
\affiliation{Osaka University, 2-6 Yamadaoka, Suita 565-0871, Japan}
\author{Alexander Pelka}
\affiliation{Helmholtz-Zentrum Dresden-Rossendorf, Bautzner Landstra\ss e 400, 01328, Dresden, Germany}
\author{Irene Prencipe}
\affiliation{Helmholtz-Zentrum Dresden-Rossendorf, Bautzner Landstra\ss e 400, 01328, Dresden, Germany}
\author{Lisa Randolph}
\affiliation{Universit\"at Siegen, Department Physik, Walter-Flex-Stra\ss e 3, 57072 Siegen, Germany}
\author{Melanie R\"odel}
\affiliation{Helmholtz-Zentrum Dresden-Rossendorf, Bautzner Landstra\ss e 400, 01328, Dresden, Germany}
\affiliation{Technische Universität Dresden, 01069 Dresden, Germany}
\author{Youichi Sakawa}
\affiliation{Osaka University, 2-6 Yamadaoka, Suita 565-0871, Japan}
\author{Hans-Peter Schlenvoigt}
\affiliation{Helmholtz-Zentrum Dresden-Rossendorf, Bautzner Landstra\ss e 400, 01328, Dresden, Germany}
\author{Michal Šmíd}
\affiliation{Helmholtz-Zentrum Dresden-Rossendorf, Bautzner Landstra\ss e 400, 01328, Dresden, Germany}
\author{Franziska Treffert}
\affiliation{SLAC National Accelerator Laboratory, 2575 Sand Hill Rd, Menlo Park, CA 94025, USA}
\affiliation{Technische Universität Darmstadt, Karolinenpl. 5, 64289 Darmstadt, Germany}
\author{Katja Voigt}
\affiliation{Helmholtz-Zentrum Dresden-Rossendorf, Bautzner Landstra\ss e 400, 01328, Dresden, Germany}
\affiliation{Technische Universität Dresden, 01069 Dresden, Germany}
\author{Karl Zeil}
\affiliation{Helmholtz-Zentrum Dresden-Rossendorf, Bautzner Landstra\ss e 400, 01328, Dresden, Germany}
\author{Thomas E. Cowan}
\author{Ulrich Schramm}
\affiliation{Helmholtz-Zentrum Dresden-Rossendorf, Bautzner Landstra\ss e 400, 01328, Dresden, Germany}
\affiliation{Technische Universität Dresden, 01069 Dresden, Germany}
\author{Thomas Kluge}
\affiliation{Helmholtz-Zentrum Dresden-Rossendorf, Bautzner Landstra\ss e 400, 01328, Dresden, Germany}
\date{\today}










\begin{abstract}

Extreme states of matter exist throughout the universe e.g. inside planetary cores, stars or astrophysical jets. 
Such conditions are generated in the laboratory in the interaction of powerful lasers with solids, and their evolution can be probed with femtosecond precision using ultra-short X-ray pulses to study laboratory astrophysics, laser-fusion research or compact particle acceleration. 
X-ray scattering (SAXS) patterns and their asymmetries occurring at X-ray energies of atomic bound-bound transitions contain information on the volumetric nanoscopic distribution of density, ionization and temperature. 
Buried heavy ion structures in high intensity laser irradiated solids expand on the nanometer scale following heat diffusion, and are heated to more than $2$ million Kelvin. 
These experiments demonstrate resonant SAXS with the aim to better characterize dynamic processes in extreme laboratory plasmas. \\


\end{abstract}
\maketitle 


High energy and high intensity (HI) lasers can compress and heat solids to extreme states of warm dense matter (WDM) and high energy density plasmas (HEDP), important for planetary science\cite{Kraus2017} and astrophysics\cite{Remington1999,*Bulanov2015a}, fusion energy research\cite{LePape2018} as well as the investigation of material properties\cite{McBride2019} and radiative properties of plasmas\cite{Bailey2015}. 
Understanding and controlling fundamental interaction processes of HI lasers and solids such as absorption at the solid-density surfaces, ultrafast growth of instabilities or two-stream unstable electron transport require measurements with high spatial and temporal precision\cite{Ge2013,Sgattoni2015,Chawla2013,*Gode2017,*Leblanc2014}. \\
Samples with a grated front surface, for example, have shown complete laser absorption resulting in enhanced ion acceleration and extreme ultra-violet (XUV) generation\cite{Kahaly2008,andreev2011efficient}. 
Multilayer structures containing few hundred nanometers to micron deep buried structures with nanoscale spatial modulation are ideal for studies of isochoric or shock heating of solid-density plasmas\cite{Huang2013,*Kluge2016}, or can serve as an ex-situ model of DT layered fusion capsules to study the role of surface roughness and modulations by adding a grating layer. 
Moreover, probing the nanoscopic evolution of structures generated during and after the laser irradiation provides a new opportunity to test and develop our fundamental understanding of 3D effects, such as volumetric\cite{Metzkes2014,Leblanc2014} or surface instabilities\cite{Kluge2015,*Palmer2012,*Macchi2001,*Sentoku2000} and their dependence on the ultrafast ionization dynamics in solid-density plasmas\cite{Huang2016}, which play a critical role in determining the quality of laser-accelerated proton and ion beams. 
This is highly relevant for the development of next-generation laser plasma based particle accelerators, where particle beam quality will be essential, for example in the context of fast ignition\cite{Jarrott2016,Azechi2013} or radiation therapy\cite{Masood2017}.\\
\begin{figure*}
\centering
      \includegraphics[width=0.96\textwidth]{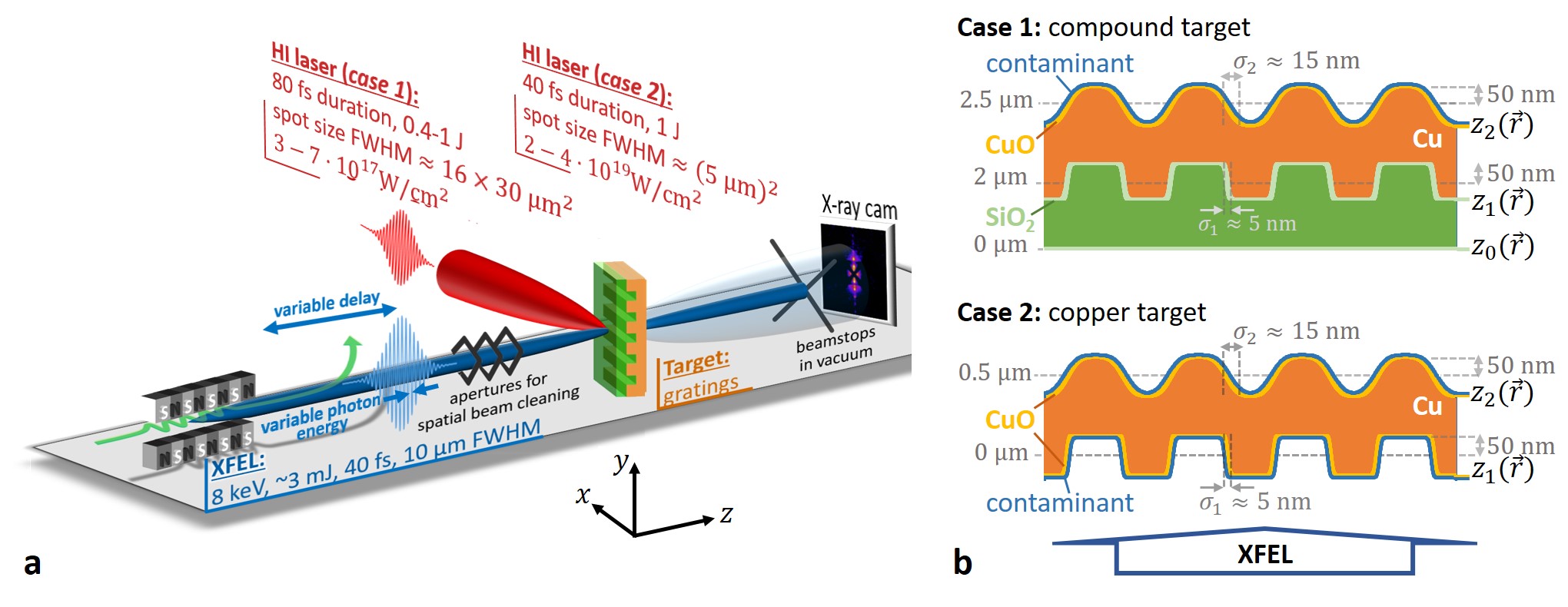}
\caption{Schematic of the Resonant-SAXS experimental setup (\textbf{a}) and target geometry (\textbf{b}) (not to scale). A near infrared HI laser pulse is focused onto a silicon-copper compound target with a grating interface (\textit{case 1}) or copper grating (\textit{case 2}). The profile edges are assumed to follow an error function, $z\propto \mathrm{erf}\left[\left(y-y_0\right)/\left(\sqrt{2}\sigma_{1/2}\right)\right]$. Both the HI drive laser  and XFEL probe beam axes are oriented at $45^\circ$ w.r.t. the target surface normal, parallel to the grating ridges and polarized as shown. . 
}
\label{fig:setup}
\end{figure*}
In order to measure the above mentioned processes and develop HI laser-based applications, as well as to open up new possibilities for fundamental relativistic plasma physics by enabling direct comparison to models and simulations, it is necessary to probe the plasma 
on the fundamental scales of down to a few nanometers and femtoseconds. 
The only source that has been shown to achieve this in solid-density plasmas currently are X-ray free-electron lasers (XFELs) \cite{Emma2010,Pile2011,Tschentscher2017}. 
The ultrashort pulses of hard X-rays produced by XFELs provide the necessary penetration power, and high spatial and temporal resolution for pump-probe experiments of HI laser-solid interaction processes.
The complex plasma dynamics becomes accessible via the temporal evolution of the nanoscale Fourier components of the scattering length density $\rho(\vec{r},t)=n_e + n_i f_i$, where $n_e$ and $n_i$ are the free electron and ion densities, respectively, and $f_i$ is the ion form factor. 
The use of structured targets has proven to be particularly useful and sensitive to nanoscopic, ultrafast dynamics employing small-angle X-ray scattering (SAXS)\cite{Gorkhover2016,Kluge2018}. \\
However, the Fourier components from structures at different depths superimpose on the detector as the complex-valued scattering length density distribution is integrated along the beam direction. 
This hinders the disentanglement of the temporal density evolution of the different components so that single-pulse SAXS experiments so far give access to the 2D areal density distribution only. \\
Here we show on the examples of grating and buried grating targets that this drawback can be overcome and 3D nanoscopic information on density and even opacity correlations can be obtained by evaluating the asymmetry $\eta(\vec q)=\left(I(\vec{q})-I(-\vec{q})\right)/\left(I(\vec{q})+I(-\vec{q})\right) \label{eqn:friedel}$ in a SAXS intensity pattern $I(\vec{q})$, where $\vec{q}$ is a vector in the reciprocal space. 
Specifically, our experiments constitute the first ultrafast measurements of the solid-density opacity using SAXS patterns. 
This method is especially interesting as in the future it could provide for opacity measurements and therefore ionization state measurements with the SAXS resolution down to nanometer scales\cite{Kluge2016}, similar to the element-specific imaging technique in\cite{Song2008}. 

\section{Asymmetry in the scattering patterns}
SAXS patterns, being proportional to the absolute square of the Fourier transform of the scattering length density $\rho$, are usually symmetric since $\rho$ is often either purely real-valued, purely imaginary or the real and imaginary parts are proportional to each other. 
Hence, any asymmetry can be directly linked to the different distribution of the real and imaginary parts of $\rho$ in the target. 
In the case of hard X-rays, the real part is primarily due to Thomson scattering, while the imaginary part arises from the opacity, i.e. the combined absorption cross section of atomic resonant bound-bound (bb), bound-free (bf), and free-free (ff) transitions. 
In the scattering length density the opacity enters via optical corrections $f'_i+i f''_i$ that need to be added to the ion form factor $f_i$. \\
We demonstrate that the asymmetry arising from differences in the distributions of electrons and ions can be used to reconstruct spatial correlations between 3D structures and their respective material properties. 
Here, we use this information for the investigation of the ultrafast plasma density expansion and non-equilibrium ionization or excitation of ions after HI laser irradiation.

Both the grating and buried grating targets we employed in our experiments have two grated interfaces (see Fig.~\ref{fig:setup}). 
We can express the two structured contours of the front (or buried) and rear surface interface as ${z}_1(\vec{r},t)$ and ${z}_2(\vec{r},t)$, respectively. 
Correlations between both contours are expressed in reciprocal space by a replication factor $\chi(\vec{q},t)$ with
\begin{equation}
\chi(\vec{q},t)=\frac{\tilde{z}_2(\vec{q},t)}{\tilde{z}_1(\vec{q},t)}.
\label{eqn:replicationFactor}
\end{equation} 
A replication factor of $\left|\chi(\vec{q},t)\right|=1$ means perfect replication, i.e. the two interfaces have identical profiles. 

In the Supplementary, Sec. IID we show how to calculate the asymmetries for the target configurations relevant for our experiments. 
The main finding is that for two-layer targets the replication factor $\chi\text{\small{$(\vec{q})$}}$ and the respective optical properties of the target materials can be determined from the asymmetry in the scattering patterns via 
\begin{equation}
    \eta\left(\vec q,t\right)=\frac{c_1\Im{\left(\chi\left(\vec{q},t\right)\right)}}{\Re{\left(c_2+c_3\,\chi\text{\small{$(\vec q,t)$}}^2+c_4\,\chi\text{\small{$(\vec q,t)$}}\right)}} 
    \label{eqn:eta_final}
\end{equation}
if the phase of the XFEL is plane\footnote{This assumption is valid for our experiments since as shown in the supplementary the preshots of the same target show similar asymmetry to each other. Therefore the effect of the phase on the asymmetry, if present at all, can be assumed to be constant, i.e. all changes of the asymmetry between pre- and main shots cannot be attributed to the influence of a phase, which will therefore for reasons of simplicity be neglected throughout the paper.}. 
This equation connects the asymmetry with the structural parameters (entering in $\chi$) and the material properties such as density and opacity (entering in $c_i$).
In experimental conditions, surface inhomogeneities (i.e.random  variations in the geometric parameters) and contamination layers  of oxides and hydrocarbon compounds are unavoidable. Their influence can be absorbed into the parameters $c_i$ and are taken into account in the calculations for the expected asymmetries of our targets which are used later in figures~\ref{fig:eta_final} and~\ref{fig:sim2main}f.\\
Before we present our results regarding the opacity measurements, we need to study the role of the replication factor. 
In fact, we will demonstrate that the asymmetry can be used to obtain nanoscopic information about interface structures lying behind each other.
We performed dedicated experiments for both cases: \textit{Case 1} is an experiment measuring the relative expansion dynamics of interface structures placed at different depths. 
Here, the laser pump intensity was chosen small enough that only structural changes occur, i.e. only the replication factor $\chi$ varies while the optical properties of the compound materials remain constant. In \textit{case 2} we focus on changes in the asymmetry caused by variations of the optical properties of the target material (via $c_i$). 
Here, a combination of high pump laser intensity and short time scales yield a rapid heating, ionization and consequently an increase of X-ray opacity at resonant bb transitions. \\

\begin{figure}[h]
\centering
      \includegraphics[width=0.6\linewidth]{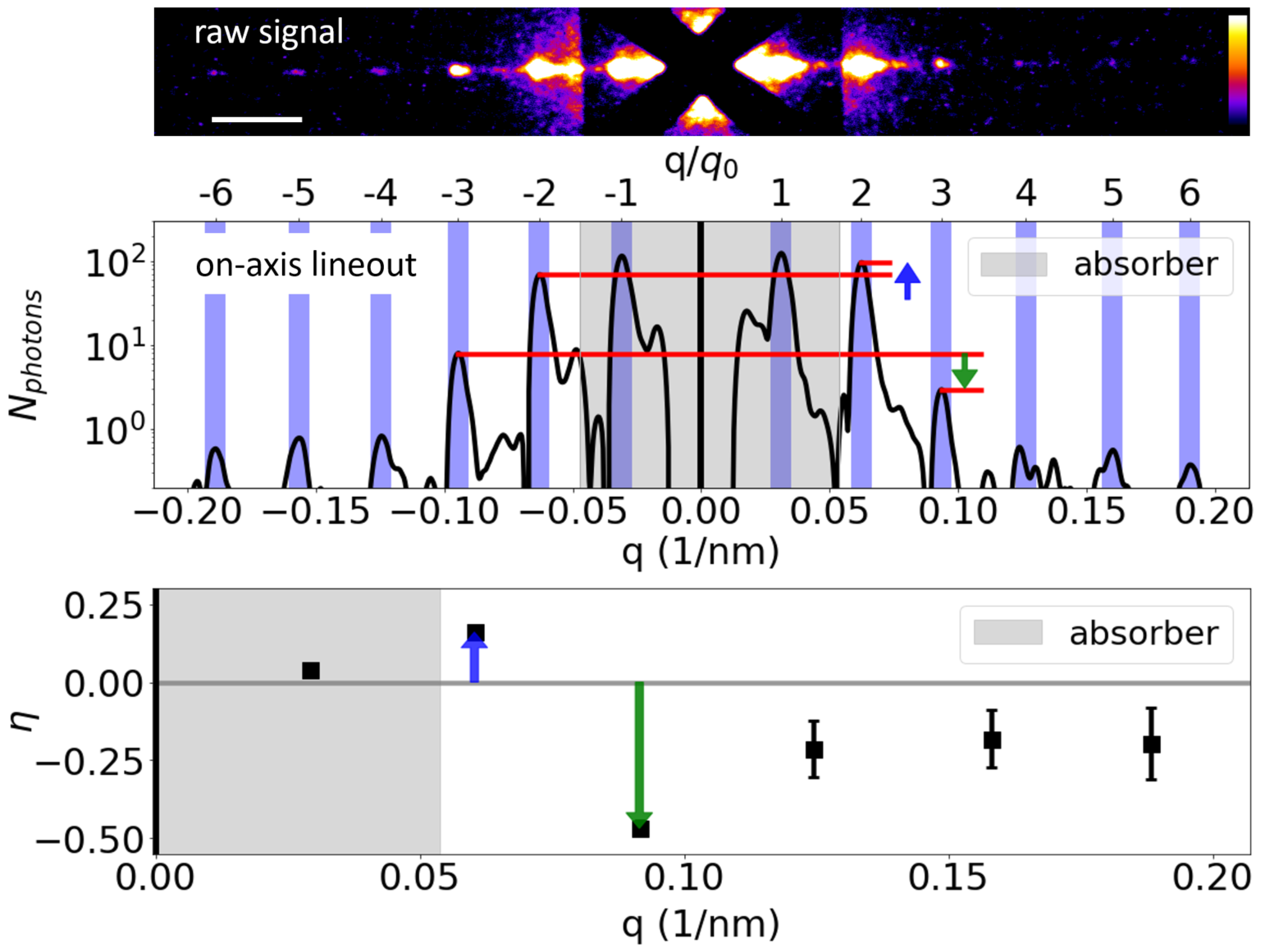}
\caption{Data processing steps from raw signal to asymmetry on the example of a resonantly probed HI laser driven Cu grating foil. \textbf{Top}: raw data (log color scale), white bar corresponds to $0.03\unit{nm}^{-1}$; \textbf{middle}: background corrected lineout through center, blue bars indicate the q-ranges for integration of the photon number per peak; \textbf{bottom}: asymmetry of peak pairs whose averaged absolute values give the average asymmetry $\overline{\left|\eta\right|}$ (see Methods). }
\label{fig:procedure}
\end{figure}

Experimentally, we extract the asymmetry by comparing the scattering signal in the scattering peaks for each peak pair at $+q$ and $-q$. 
The absolutes of the individual asymmetry values are then averaged in order to have a simple quantity to easily compare simulations with theory. 
Additionally, the uncertainty margins of the averages are much smaller than those of the individual asymmetry values. 
Fig.~\ref{fig:procedure} illustrates the workflow from the raw experimental data to the average asymmetry values $\overline{\left|\eta\right|}$. 

\section{Case 1: Geometric changes}
For this part of the experiment, Si-Cu compound targets were pumped at an intensity of 3 and 7$\times 10^{17} \unit{\mathrm{W}/\mathrm{cm}}^2$ of the HI laser (see Fig.\ref{fig:setup}). 
At this comparatively low intensity, the resulting heating is sufficiently low such that collisional ionization of the rear copper layer is suppressed to less than the L-shell, i.e. no K$\alpha$ transition channels are available. 
Additionally, a slow effective expansion of either one or both the buried grating interface and rear grated surface (i.e. an increase of $\sigma_{1}$ or $\sigma_{2}$) can be deduced from a decrease of the scattering yield at large values of $q$ with a Debye-Waller-like procedure similar to that presented in~\cite{Kluge2018}.
Comparing the SAXS patterns obtained after irradiation by the HI laser pulse (main shots) with those obtained before from the target without HI irradiation  (preshots), we observe a reduction of scattering yield at large $q$-values corresponding very roughly by an effective expansion of up to $10\unit{nm}$ (see Supplementary). 
However, the expansion values cannot be fitted separately for the interface and rear surface with a reasonable accuracy since the projected total electron density of the target with two grating surfaces simply has too many free parameters.\\
On the other hand, no significant reduction of the yield was found for the shortest probe delay of $250\unit{fs}$, indicating that the interface expansion indeed is a slow process over a few ps, as expected for those low laser intensities. 

\begin{figure}
\centering
      \includegraphics[width=\linewidth]{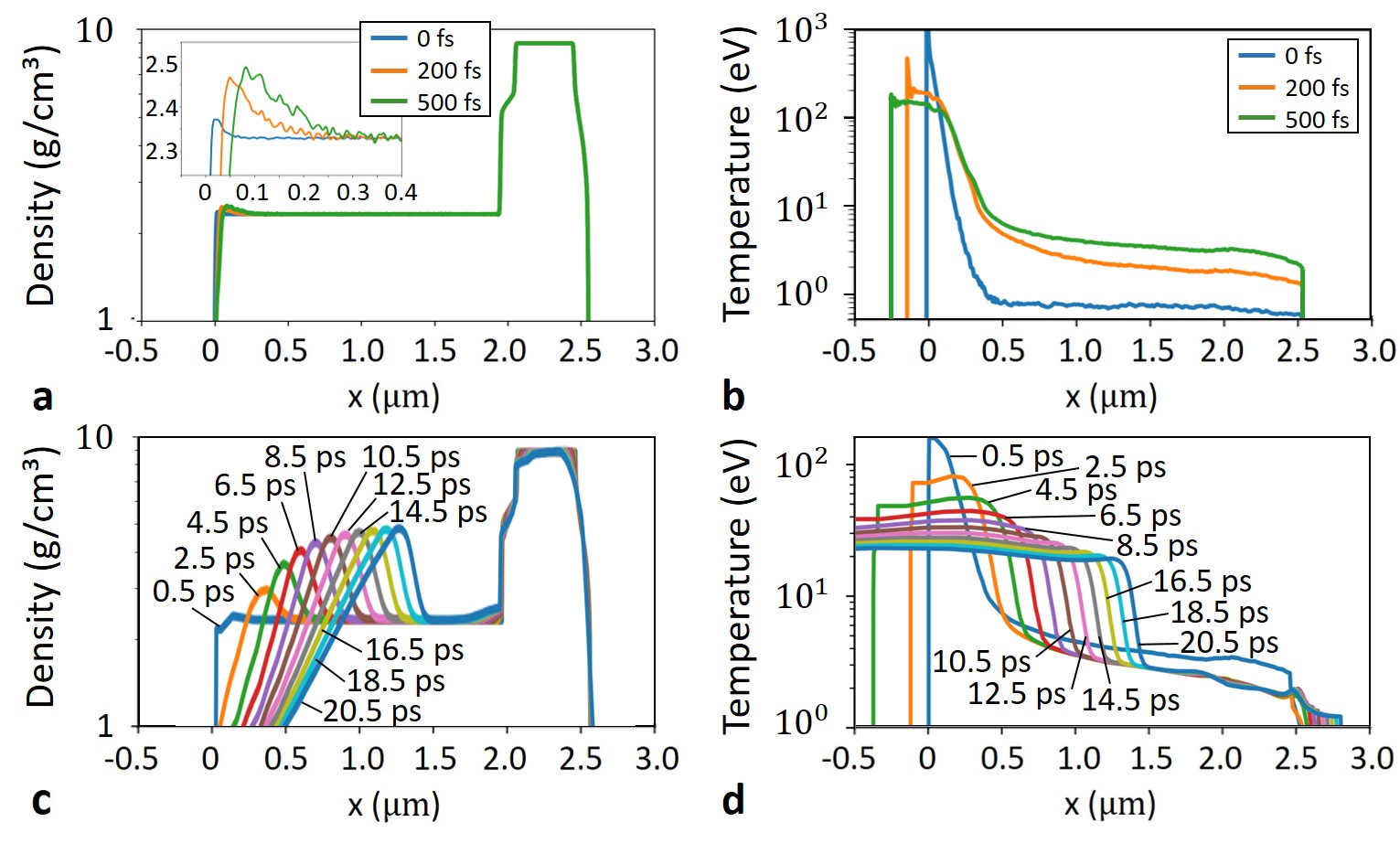}
\caption{Density (\textbf{a},\textbf{c}) and temperature (\textbf{b},\textbf{d}) evolution of the Si-Cu compound target for the average optical laser intensity in the XFEL probe area from Particle in cell simulations (\textbf{a},\textbf{b}) and hydrodynamic simulations (\textbf{c},\textbf{d}). Laser strength was set to $a_0\cong 0.3$, which corresponds to the average intensity in the XFEL focal spot for $1\unit{J}$ laser energy. Laser is incident from the left.}
\label{fig:sim1main}
\end{figure}

To confirm the expectations of low heating and slow expansion, we performed particle-in-cell (PIC) simulations to model the laser interaction and prompt electron heating in the first $500\unit{fs}$ after the laser hit the target, followed by a hydrodynamic simulation to study the plasma response over the few ps range up to the $16\unit{ps}$ delay of the latest probe, see Fig.~\ref{fig:sim1main}. 
While at the front surface the high electron temperature generates a strong ablation pressure which leads to shock formation and ablation, heat diffusion towards the target rear leads to a partial temperature equilibration over the whole compound target depth to $~5\eV$ within the following few hundred femtoseconds. 
The key finding here is that the PIC simulation predicts a very slow expansion of the buried interface and the rear surface, which is in fact below the simulation's resolution within the first $500\unit{fs}$. 
With the hydrodynamic simulations we then observe the formation of a strong shock over the next $5\unit{ps}$ that propagates toward the target rear with an initial velocity of $169\unit{nm/ps}$, slowing down to $45.5\unit{nm/ps}$ after $20\unit{ps}$. 
This means that the rear side interface remains intact during the whole range of our probing delays and scattering patterns are changing only due to the thermal expansion of the grating surfaces.

\begin{figure}
      \includegraphics[width=0.7\linewidth]{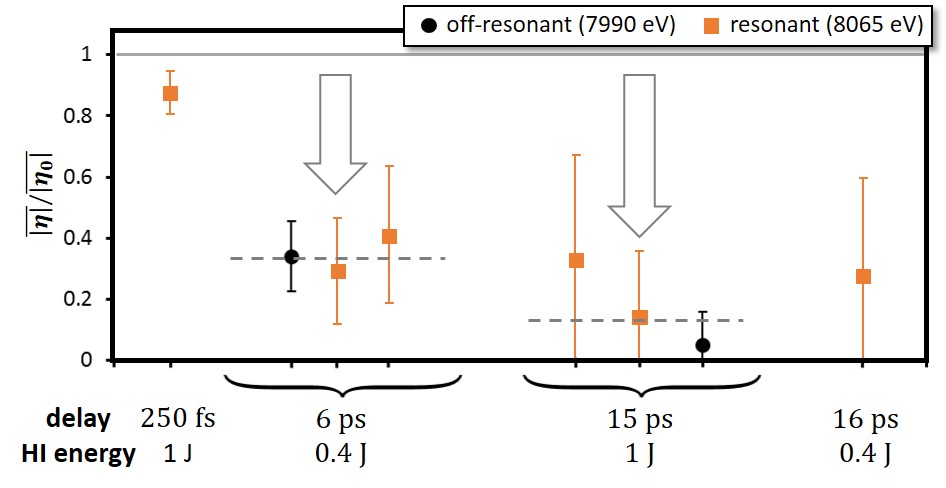}
\caption{\label{fig:asymmetries}\textit{(Case 1)} Average asymmetry $\overline{\left|\eta\right|}$ relative to that of the preshots $\overline{\left|\eta_0\right|}$ for single shots on Si-Cu compound targets with $3-7 \times 10^{17}\unit{W/cm}^2$ laser intensity. Dashed lines are respective weighted averages. }
\end{figure}
We now analyze the asymmetry parameter $\overline{\left|\eta\right|}$ in the experimental scattering patterns in order to obtain additional information about the relative expansion of both gratings. 
We observe a reduction of $\overline{\left|\eta\right|}$ compared to its value $\overline{\left|\eta_0\right|}$ in the respective XFEL-only preshots, which is dependent on a combination of the delay and energy of the HI laser (see Fig.~\ref{fig:asymmetries}). 
As expected, since no K$\alpha$ channels are available, the reduction of the asymmetry is not substantially different for the different XFEL photon energies (i.e. on- and off-resonance). 
The observed reduction of the asymmetry must therefore be predominantly due to a change of the replication factor, i.e. expansion of the grating surfaces. 

In Fig.~\ref{fig:eta_final} the calculated asymmetries (see Supplementary, Sec. IID) are shown as a function of the two respective smoothness parameters $\sigma_1$ and $\sigma_2$ of the buried and rear grating ridges. 
We adopt a replication factor of $\chi\small{\left(\vec{q}\right)} = \hat \chi \exp{\left(-\Delta\sigma^2 q^2/2\right)}$ with $\Delta\sigma^2\equiv \sigma_2^2-\sigma_1^2$, i.e. a smoothing of the rear interface (index 2) with respect to the buried interface (index 1). 
With the initial conditions for the un-pumped targets (see Methods), it can be readily seen that a reduction in asymmetry is linked to an expansion of the grating interfaces. 
Specifically, the experimentally observed reduction by more than a factor of three reduces the range of possible values for $\sigma_1$ and $\sigma_2$. 
The result we can therefore infer is that the observed drop in asymmetry is characteristic for the expansion of the buried grating interface ($\sigma_1$) by at least $10\unit{nm}$ and cannot be explained by an expansion of the rear grating interface ($\sigma_2$) alone. 

\begin{figure}
\centering
      \includegraphics[width=0.6\linewidth]{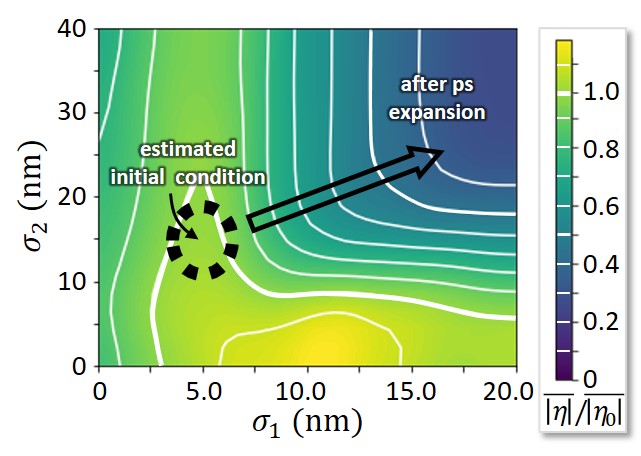}
\caption{Analytical modeling of the asymmetry $\left|\overline{\eta}\right|$ averaged over the experimentally accessible $q$-range and normalized by the corresponding preshot asymmetry $\overline{\left|\eta_0\right|}$ for \textit{case 1} (Si-Cu compound targets). }
\label{fig:eta_final}
\end{figure}

\section{Case 2: Opacity changes}
\begin{figure}
      \includegraphics[width=0.65\linewidth]{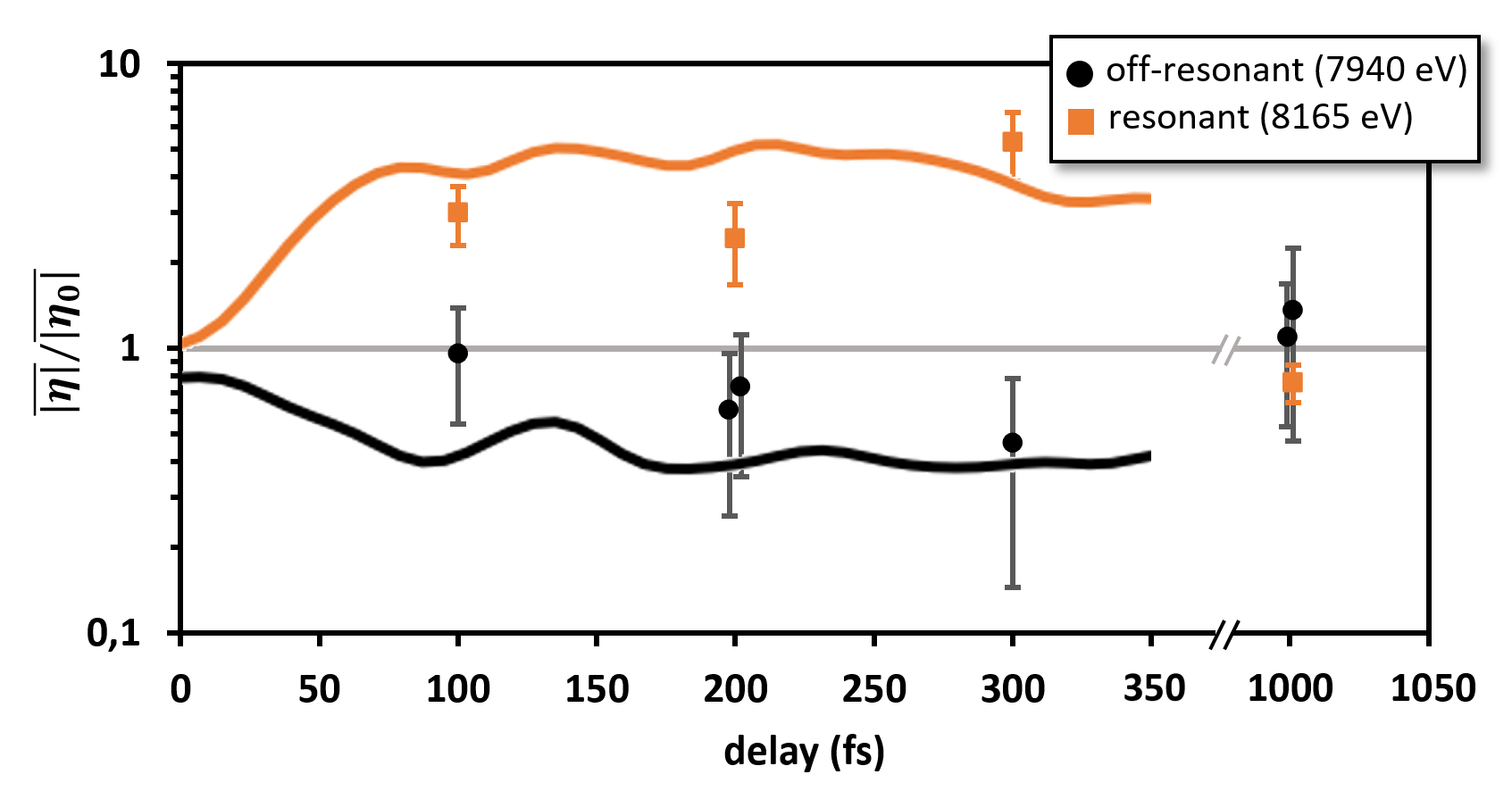}
\caption{\label{fig:LT94}\textit{(Case 2)} Average asymmetries $\overline{\left|\eta\right|}$ relative to that of the preshots $\overline{\left|\eta_0\right|}$ for the shots on Cu targets with $2-4 \times 10^{19}\unit{W/cm}^2$ laser intensity. Solid lines show asymmetry obtained from PIC simulations for off-resonant (black) and resonant (orange) XFEL energy. 
} 

\end{figure}

The second case considered in this paper focuses on  ionization-induced changes in the asymmetry of the SAXS signal as a novel tool to measure the temporal evolution of plasma opacities. 
For this purpose, we used Cu grating targets irradiated by a higher laser intensity with a peak intensity of $(3\pm1)\times 10^{19}\unit{\mathrm{W}/\mathrm{cm}}^2$. 
With the first grating positioned now directly at the front surface, the dynamics are no longer dominated by heat diffusion and shock propagation through the silicon layer as before in \textit{case 1}, but by the faster direct laser-solid interaction and by hot electrons accelerated into the target by the laser.
The relevant timescales for these processes are shorter, on the order of a few hundred femtoseconds only. 
Consequently, we start to observe a slight decrease of the asymmetry due to the grating expansion for shots with XFEL probe delay already at $200\unit{fs}$ at off-resonant probe energies of $7940\unit{\eV}$, see Fig.~\ref{fig:LT94}. 

For resonant scattering, we tuned the energy of the XFEL to an open atomic transition in the plasma, namely to $8165\unit{eV}$. 
This is the energy of the K$\alpha$ transition in nitrogen-like Cu (K1L6 - K2L5)\cite{Smid2019}. 
During the first $300\unit{fs}$ after pump laser irradiation the scattering patterns in those resonant shots exhibited large values of asymmetry compared to the XFEL-only preshots, cf. Fig.~\ref{fig:LT94}.
The increase in asymmetry only disappears at larger delays. 
At the largest delay of $1\unit{ps}$ the main shots and preshots show a similar asymmetry again. 
This is to be expected, due to the onset of expansion/destruction of the front side grating and ion-electron recombination (see below). 

In order to understand these observations, we refer to the discussion in the previous \textit{case 1}.
We expect the rear side grating to remain virtually unchanged while the front surface should thermally expand during the first few hundred femtoseconds after the laser irradiation, which is also supported by PIC simulations described later. 
Consequently, the reduced asymmetry in the main shots at off-resonant XFEL energy during the initial $300\unit{fs}$ after the HI irradiation compared to the preshots can be explained by a reduced replication factor $\chi$. 
On the other hand, this means that the increase of asymmetry for the on-resonance shots at the same delays can not be caused by geometric changes and hence must be due to a change of the opacity at the resonant energy.
Since the cross section of bf transitions is similar within the range of the two XFEL photon energies used in the experiment (within $\pm 25\%$), the large increase of asymmetry in the on-resonance shots must be due to an increase of $f''_{Cu}$ due to bb transitions\cite{Kluge2017}, whose existence was measured independently with a temporally and spatially integrating spectrometer (see Fig.~2 in Methods). 

\begin{figure*}
\centering
      \includegraphics[width=\linewidth]{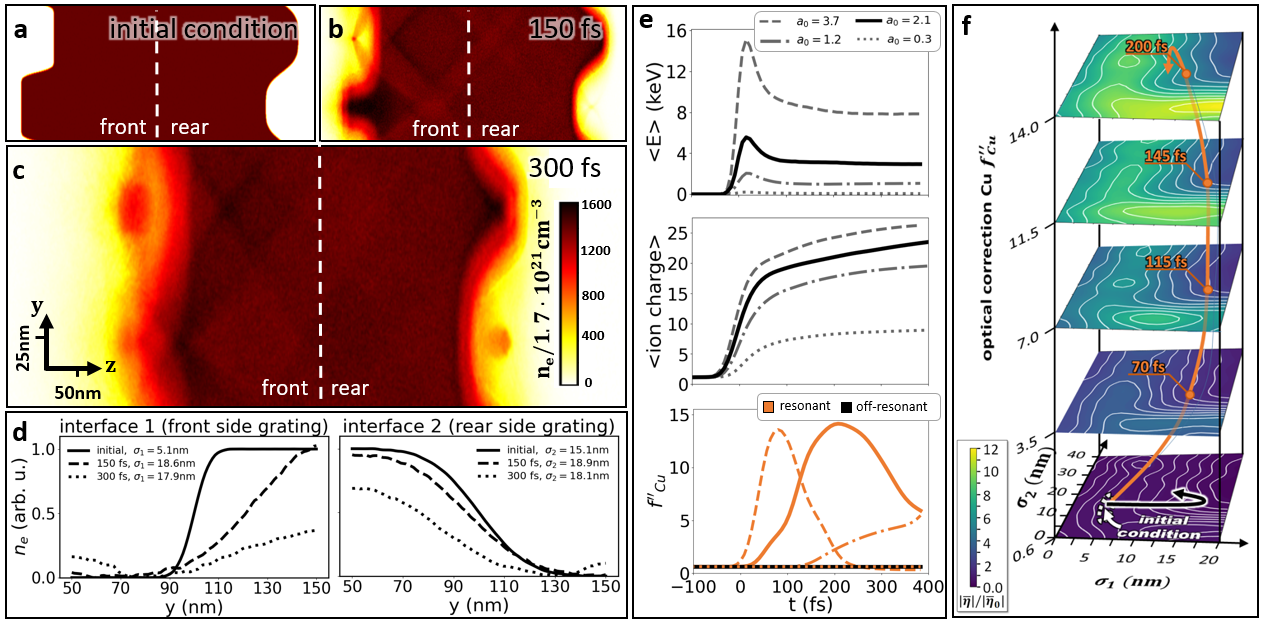}
\caption{Simulation of the Cu target (\textit{case 2}) for the average intensity of $1.3\cdot 10^{19} \unit{W/cm}^2$ ($a_0=2.1$). \textbf{a-c} Total electron density distribution at three different times. \textbf{d} Transverse electron density integrated through the first (left) and the rear half of the target (right) normalized with respect to the initial grating amplitude. \textbf{e} Temporal evolution of the average electron energy (top), average Copper ionization charge state (middle), and fitted opacity using SCFLY\cite{scfly} (bottom) for resonant (orange, $8165\unit{eV}$) and off-resonant (black, $7940\unit{eV}$) XFEL energy. For reference the evolution is given also for different laser intensities, corresponding do different parts of the laser focal spot. \textbf{f} Calculated asymmetry based on the simulation results for resonant (orange) and off-resonance (black) XFEL energy.}
\label{fig:sim2main}
\end{figure*}


We therefore conclude that the apparent plasma temperature is above $300\unit{eV}$, which is the necessary temperature to sufficiently ionize copper to open the nitrogen-like K$\alpha$ resonances at the probe photon energy of $8165\unit{eV}$\cite{Smid2019,scfly}. 
However, the plasma evolution is more complex, i.e. non-thermal and highly transient, as the laser-plasma interaction is pulsed, relativistic and non-linear. 
In order to model the plasma evolution, we performed a set of PIC simulations. 
We varied the relativistic laser amplitude $a_0$ to model different regions inside the UHI laser focal spot. 

During the first few tens to hundreds of femtoseconds following the HI laser irradiation, the grating is quickly ionized and electrons are accelerated and heated. 
The simulations for $a_0=3.7$ down to $a_0=1.2$ show average free electron energies in the range of multiple keV; for the nominal average intensity in the grating area, i.e. $a_0=2.1$, it peaks at approximately $\langle E\rangle =5\unit{\keV}$, cf. Fig.~\ref{fig:sim2main}.  
It is worth mentioning that in thermal equilibrium an electron temperature in this range would suggest that Cu should be ionized up to He- and H-like states. 
Consequently, we could not expect to observe N-like K$\alpha$ transitions. 
However, due to the non-equilibrium transient situation and limited ionization rates, the  average ionization can be seen and is known\cite{Huang2017} to be much less during the first several hundred femtoseconds. 
Thus, the ionization distribution contains significant contribution of N-like ionized Cu  and the respective K$\alpha$ resonance in fact \textit{can} be driven by the XFEL.


In order to estimate the increase of absorption and asymmetry, we fit the average charge state of a 0-dimensional atomic simulation (using SCFLY\cite{scfly}, varying the NLTE temperature) to that of the PIC simulation for each time step. 
From this we can then extract the opacity for the respective matched cases and consequently estimate the temporal evolution of $f''_{Cu}$ (see Fig.~\ref{fig:sim2main}e). 
Finally, we can now calculate the expected change of the asymmetry in the SAXS scattering patterns when the XFEL energy is tuned to the nitrogen-like K$\alpha$ resonance energy of $8165\unit{eV}$. 
For specific expansions of the grating surfaces extracted from the PIC simulations we can use Eqn.~\ref{eqn:eta_final} with the appropriate extension for including surface contaminants to compute the asymmetries for the respective increased absorption, see Fig.\ref{fig:sim2main}f for specific example at the average laser intensity. 
The predictions for the experimental results have to take into account the spatial laser intensity profile. 
We do this in Fig.~\ref{fig:LT94} by first computing the asymmetric scattering amplitudes for each of the simulations with different laser intensity. 
We than calculate the total scattering signal by summing up those individual contributions, weighted by the areal fraction of the respective laser intensity in the focal spot. 
The resulting semi-analytical predictions for the asymmetries of the scattering pattern can be seen to be in excellent agreement with the experimental results.

Summarizing, the fact that the asymmetry is increased during the first $300\unit{fs}$ after HI laser irradiation is due to a combination of electron acceleration to an average energy well above $300\unit{\eV}$, moderate delayed grating expansion, and comparatively slow ionization not reaching saturation during the brief heating time before the bulk cools down again.
\section{Outlook}
The potential of XFEL-based SAXS and especially resonant SAXS to measure the spatial distribution of the electron density and plasma opacity as well as correlations both in the XFEL transverse \textit{and} longitudinal direction via the replication factor offers a unique way to characterize complex dynamic plasma processes. 
We developed an analytical model that demonstrates the connection between the optical properties of a material and surface replications in the X-ray longitudinal direction with the asymmetry in the scattering pattern. 
Here we demonstrated the application of asymmetric scattering on disentangling the expansion of a buried layer from the one at the rear, and to extract the plasma opacity by tuning the XFEL energy to the nitrogen-like bb transition of copper. 
The asymmetry changes upon laser pumping were found to be in quantitative agreement with an expansion of a grating buried $2\mum$ deep in silicon of at least $10\unit{nm}$ at low laser intensities, and a heating of a thin copper target to at least $300\unit{eV}$ at HI laser intensities exceeding $10^{19}\unit{W/cm}^2$. \\
For 2D scattering patterns and stronger signals, in the future it will become possible to make use of the full $\vec{q}$-dependency of the asymmetry. 
Then, phase retrieval algorithms can be employed to directly image the complex-valued scattering amplitude\cite{Fienup:90,Leshem2016,Tanyag2015}, and to obtain the electron \textit{and} ionization state spatial distribution separately and model-free in a single shot with a single detector. 
Additionally, advanced imaging methods such as Fourier transform holography\cite{Stadler2008} and two-color or X-ray-pulse split-and-delay probing\cite{Gunther2011} can in the future help to employ the asymmetric scattering for extraction of these parameters.\\
This technique has applications in a wide range of dynamical phenomena such as laser ablation, laser heating and ionization, shock formation in warm dense matter, plasma expansion or plasma instabilities. 
It may enable the exploration of the opacity on the nanometer and femtosecond scale in non-thermal conditions, which can be used as an ultrafast high resolution thermometer for warm and hot dense matter. 
Moreover, with access to longitudinal correlations in transmission geometry and high temporal resolution, it can assist or replace tomographic methods to gain a 3D understanding of the sample structure in highly dynamical systems. 
\section*{Data availability}
The raw data used for this publication is available under GNU Lesser General Public License 3.0\cite{dataLT94Asymmetry}. 
%

\section*{acknowledgments}
This work was partially supported by DOE Office of Science, Fusion Energy Science under FWP 100182. Use of the Linac Coherent Light Source (LCLS), SLAC National Accelerator Laboratory, is supported by the U.S. Department of Energy, Office of Science, Office of Basic Energy Sciences under Contract No. DE-AC02-76SF00515. The experiments were performed at the Matter at Extreme Conditions (MEC) instrument of LCLS, supported by the DOE Office of Science, Fusion Energy Science under contract No. SF00515. 
This work has also been supported by HIBEF (www.hibef.eu) and partially by Horizon 2020 LASERLAB-EUROPE (Contract No.  871124) and by the German Federal Ministry of Education and Research (BMBF) under contract number 03Z1O511. 
This work was partially  funded by the Center of Advanced Systems Understanding (CASUS), which is financed by Germany’s Federal Ministry of Education and Research (BMBF) and by the Saxon Ministry for Science, Culture and Tourism (SMWKT) with tax funds on the basis of the budget approved by the Saxon State Parliament.\\
The results of the  Project LQ1606 were obtained with the financial support of the Czech Ministry of Education, Youth and Sports as part of targeted support from the National Programme of Sustainability  II. Supported by CAAS project CZ.02.1.01/0.0/0.0/16\_019/0000778 from European Regional Development Fund; Czech Technical University grant SGS16/247/OHK4/3T/14 and Czech Science Foundation project 18-20962S. The computations were performed using computational resources funded from the CAAS project. This work has received funding from the Eurofusion Enabling Research Project No. ENR-IFE19.CEA-01.\\
CG acknowledges funding via DFG GU 535/6-1.\\

\section{Supplementary Materials:}
\subsection{Materials and Methods}
\subsection{Supplementary text}

\end{document}


\title{Supplementary Materials for: Probing ultrafast laser plasma processes inside solids with resonant small-angle X-ray scattering} 

\author{Lennart Gaus}
\affiliation{Helmholtz-Zentrum Dresden-Rossendorf, Bautzner Landstra\ss e 400, 01328, Dresden, Germany}
\affiliation{Technische Universität Dresden, 01069 Dresden, Germany}
\author{Lothar Bischoff}
\affiliation{Helmholtz-Zentrum Dresden-Rossendorf, Bautzner Landstra\ss e 400, 01328, Dresden, Germany}
\author{Michael Bussmann}
\affiliation{Helmholtz-Zentrum Dresden-Rossendorf, Bautzner Landstra\ss e 400, 01328, Dresden, Germany}
\affiliation{Center for Advanced Systems Understanding (CASUS), Görlitz, Germany}
\author{Eric Cunningham}
\affiliation{SLAC National Accelerator Laboratory, 2575 Sand Hill Rd, Menlo Park, CA 94025, USA}
\author{Chandra B.  Curry}
\affiliation{SLAC National Accelerator Laboratory, 2575 Sand Hill Rd, Menlo Park, CA 94025, USA}
\affiliation{University of Alberta, 116 St. and 85 Ave. Edmonton, AB T6G 2R3, Canada}
\author{Eric Galtier}
\affiliation{SLAC National Accelerator Laboratory, 2575 Sand Hill Rd, Menlo Park, CA 94025, USA}
\author{Maxence Gauthier}
\affiliation{SLAC National Accelerator Laboratory, 2575 Sand Hill Rd, Menlo Park, CA 94025, USA}
\author{Alejandro Laso Garc\'ia}
\affiliation{Helmholtz-Zentrum Dresden-Rossendorf, Bautzner Landstra\ss e 400, 01328, Dresden, Germany}
\author{Marco Garten}
\affiliation{Helmholtz-Zentrum Dresden-Rossendorf, Bautzner Landstra\ss e 400, 01328, Dresden, Germany}
\affiliation{Technische Universität Dresden, 01069 Dresden, Germany}
\author{Siegfried Glenzer}
\affiliation{SLAC National Accelerator Laboratory, 2575 Sand Hill Rd, Menlo Park, CA 94025, USA}
\author{J\"org Grenzer}
\affiliation{Helmholtz-Zentrum Dresden-Rossendorf, Bautzner Landstra\ss e 400, 01328, Dresden, Germany}
\author{Christian Gutt}
\affiliation{Universit\"at Siegen, Department Physik, Walter-Flex-Stra\ss e 3, 57072 Siegen, Germany}
\author{Nicholas J. Hartley}
\affiliation{Helmholtz-Zentrum Dresden-Rossendorf, Bautzner Landstra\ss e 400, 01328, Dresden, Germany}
\affiliation{SLAC National Accelerator Laboratory, 2575 Sand Hill Rd, Menlo Park, CA 94025, USA}
\author{Lingen Huang}
\affiliation{Helmholtz-Zentrum Dresden-Rossendorf, Bautzner Landstra\ss e 400, 01328, Dresden, Germany}
\author{Uwe H\"ubner}
\affiliation{Leibniz Institute of Photonic Technology, Albert-Einstein-Stra\ss e 9, 07745 Jena, Germany}
\author{Dominik Kraus}
\affiliation{Helmholtz-Zentrum Dresden-Rossendorf, Bautzner Landstra\ss e 400, 01328, Dresden, Germany}
\author{Hae Ja Lee}
\affiliation{SLAC National Accelerator Laboratory, 2575 Sand Hill Rd, Menlo Park, CA 94025, USA}
\author{Emma E. McBride}
\affiliation{SLAC National Accelerator Laboratory, 2575 Sand Hill Rd, Menlo Park, CA 94025, USA}
\affiliation{European XFEL, Holzkoppel 4, 22869 Schenefeld, Germany}
\author{Josefine Metzkes-Ng}
\affiliation{Helmholtz-Zentrum Dresden-Rossendorf, Bautzner Landstra\ss e 400, 01328, Dresden, Germany}
\author{Bob Nagler}
\affiliation{SLAC National Accelerator Laboratory, 2575 Sand Hill Rd, Menlo Park, CA 94025, USA}
\author{Motoaki Nakatsutsumi}
\affiliation{European XFEL, Holzkoppel 4, 22869 Schenefeld, Germany}
\author{Jan Nikl}
\affiliation{ELI Beamlines Centre, Institute of Physics, Czech Academy of Sciences, Za Radnicí 835, Dolní Břežany 25241, Czech Republic}
\affiliation{Faculty of Nuclear Sciences and Physical Engineering, Czech Technical University in Prague, Břehová 7, Prague 11519, Czech Republic}
\affiliation{Institute of Plasma Physics, Czech Academy of Sciences, Za Slovankou 1782/3, Prague 18200, Czech Republic}
\author{Masato Ota}
\affiliation{Osaka University, 2-6 Yamadaoka, Suita 565-0871, Japan}
\author{Alexander Pelka}
\affiliation{Helmholtz-Zentrum Dresden-Rossendorf, Bautzner Landstra\ss e 400, 01328, Dresden, Germany}
\author{Irene Prencipe}
\affiliation{Helmholtz-Zentrum Dresden-Rossendorf, Bautzner Landstra\ss e 400, 01328, Dresden, Germany}
\author{Lisa Randolph}
\affiliation{Universit\"at Siegen, Department Physik, Walter-Flex-Stra\ss e 3, 57072 Siegen, Germany}
\author{Melanie R\"odel}
\affiliation{Helmholtz-Zentrum Dresden-Rossendorf, Bautzner Landstra\ss e 400, 01328, Dresden, Germany}
\affiliation{Technische Universität Dresden, 01069 Dresden, Germany}
\author{Youichi Sakawa}
\affiliation{Osaka University, 2-6 Yamadaoka, Suita 565-0871, Japan}
\author{Hans-Peter Schlenvoigt}
\affiliation{Helmholtz-Zentrum Dresden-Rossendorf, Bautzner Landstra\ss e 400, 01328, Dresden, Germany}
\author{Michal Šmíd}
\affiliation{Helmholtz-Zentrum Dresden-Rossendorf, Bautzner Landstra\ss e 400, 01328, Dresden, Germany}
\author{Franziska Treffert}
\affiliation{SLAC National Accelerator Laboratory, 2575 Sand Hill Rd, Menlo Park, CA 94025, USA}
\affiliation{Technische Universität Darmstadt, Karolinenpl. 5, 64289 Darmstadt, Germany}
\author{Katja Voigt}
\affiliation{Helmholtz-Zentrum Dresden-Rossendorf, Bautzner Landstra\ss e 400, 01328, Dresden, Germany}
\affiliation{Technische Universität Dresden, 01069 Dresden, Germany}
\author{Karl Zeil}
\affiliation{Helmholtz-Zentrum Dresden-Rossendorf, Bautzner Landstra\ss e 400, 01328, Dresden, Germany}
\author{Thomas E. Cowan}
\author{Ulrich Schramm}
\affiliation{Helmholtz-Zentrum Dresden-Rossendorf, Bautzner Landstra\ss e 400, 01328, Dresden, Germany}
\affiliation{Technische Universität Dresden, 01069 Dresden, Germany}
\author{Thomas Kluge}
\affiliation{Helmholtz-Zentrum Dresden-Rossendorf, Bautzner Landstra\ss e 400, 01328, Dresden, Germany}

\date{\today}

\maketitle
\textbf{This PDF file includes:}

Materials and Methods

Supplementary text

\section{Materials and methods}
\color{white}.\color{black}\\
\\
\underline{Target samples}\\
For the experiments we used grated layer targets (schematic drawing in Fig.~1b of the main text and scanning electron microscopy (SEM) images in Fig.~\ref{fig:targets} below). 
For \textit{case 1} we employed compound gratings of a $2\unit{\mum}$ thin silicon support membrane ($n_e=7\cdot 10^{23}\unit{cm}^{-3}$, $f''=0.33$) covered with a copper layer ($n_e=2.4\cdot 10^{24}\unit{cm}^{-3}$, $f''=0.63$). 
First, a grating was inscribed into the rear surface of the Si membrane and subsequently covered with copper of a few hundred nanometers thickness. 
With this the buried grating structure at the Si-Cu interface imprints also at the Cu-vacuum surface (see inset in Fig.~1a of the main text) yielding a replication factor $\chi\left(\vec{q}\right)=\hat\chi\exp{\left[-\left(\sigma_2^2-\sigma_1^2\right)q^2/2\right]}$
with $\hat\chi$ close to unity between the buried and rear side grating structures. 
Note that the grating interfaces are protected from direct laser interaction by the Si membrane. 
For \textit{case 2} this membrane was etched away almost completely, leaving only the copper foil with a sharp grating at the front and its less sharp imprint on the rear. \\
The target grating smoothness $\sigma_1$ for the buried grating of the silicon-copper compound targets and the front grating of the pure copper targets were taken from the preshots in\cite{Kluge2018}, as there we used the same grated silicon carrier membrane. 
There, the grating smoothness parameter was measured to be $5.4\unit{nm}$. \\
The rear surface smoothness $\sigma_2$ of the Cu coating in the present experiment cannot be accessed independently via SAXS, yet we can restrain the possible value of $\sigma_2$. 
It must be larger than $\sigma_1$ and based on the analytic analysis of the asymmetry, the results are consistent with a smoothness of $\sigma_2\approx 10-15\unit{nm}$ and $\hat\chi=1$, which is also supoorted by the SEM images (Fig.~\ref{fig:targets}). 
The buried grating height was set to the manufacturing specification of $100\unit{nm}$. 
\begin{figure}[h]
    \centering
    \includegraphics[width=0.45\linewidth]{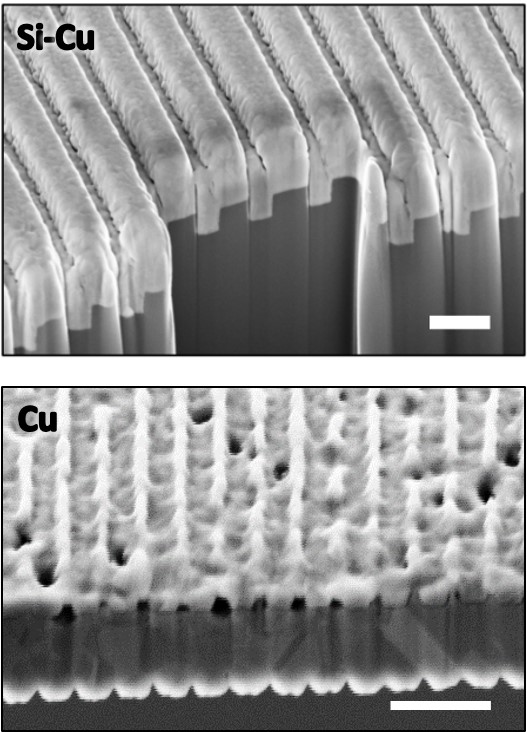}
    \caption{Scanning electron microscopy (SEM) images of representative target sample cross sections, prepared by focused ion beam cuts. \textbf{Top:} Si-Cu -- dark: Si, light gray: Cu; \textbf{Bottom:} Cu. White bar is $500\unit{nm}$}
    \label{fig:targets}
\end{figure}\\
\\
\underline{Optical laser}\\
The experiments were performed at the Matter in Extreme Conditions (MEC) endstation of the Linac Coherent Light Source (LCLS) at the Stanford Linear Accelerator Center (SLAC). 
We used the MEC short-pulse HI pump laser to generate a solid-density plasma, i.e. induce fast, non-thermal melting and ionization/excitation. 
This is a Titanium:Sapphire-based high power laser system based on chirped pulse amplification. 
The optical and XFEL configuration for \textit{case 1} is the same as described in \cite{Kluge2018}, i.e. the optical pump laser provided ultrashort pulses ($\tau = 80\unit{fs}$ at a central wavelength of $800\unit{nm}$ with an energy of 0.4\,J or 1\,J before the compressor which corresponds to 180\,mJ and 460\,mJ on target, focused to a spot size of $30\unit{\mum}\times 16\unit{\mum}$ FWHM.\\
For \textit{case 2} the laser was compressed and focused more tightly ($\tau = 40\unit{fs}$, spot size FWHM $5\unit{\mum}$) resulting in two orders of magnitude higher peak intensity on target of approximately $ 2-4\cdot 10^{19}\unit{W}/\mathrm{cm}^2$. \\
\\
\underline{XFEL}\\
The LCLS XFEL beam was used to diagnose the plasma dynamics by means of SAXS. 
We used two ranges of X-ray photon energies: off-resonant, i.e. $7990\unit{eV}$ (\textit{case 1}) and $7940\unit{eV}$ (\textit{case 2}) and resonant, i.e. $8065\unit{eV}$ (\textit{case 1}), $8165\unit{eV}$ 
(\textit{case 2}). 
At the off-resonant X-ray energies, the cross sections are dominated by photo-ionization (bf transitions) and Thomson scattering. 
Photon energies on Cu-K$\alpha$ bb resonance transitions allowed us to probe the ionization state of Cu via X-ray absorption. 
The fundamental frequency of the LCLS X-ray beam was focused with compound refractive lenses into the MEC experimental area to spot sizes of $20\unit{\mum}$ (\textit{case 1}) and between 5 and $10\unit{\mum}$ (\textit{case 2}). The $3^{rd}$ harmonic was only weakly focused and hence its intensity is significantly reduced on target. For \textit{case 2} a high harmonic rejection mirror system was used. While in \textit{case 1} the XFEL pulse intensity had to be attenuated by various Si and Cu absorbers to ensure the scattering signal was within the dynamic range of the PIXIS XF 2048B camera, for \textit{case 2} we positioned small absorber plates in front of the camera to selectively attenuate the first scattering peaks only. This allowed us to use much higher XFEL transmissions (between 20\% and 100\%) than reported in \cite{Kluge2018}. Consequently, we could measure the signal to higher values of q. \\
\\
\underline{Small-angle X-ray scattering}\\
We used a PIXIS XF 2048B X-ray camera to record the scattering pattern. For absolute photon numbers we calibrated it using an $\mathrm{Am}^{241}$ and an $\mathrm{Fe}^{55}$ source. 
The system resolution is dictated by the PIXIS point spread function and XFEL beam divergence, which are both between 2 and 3 pixels on the detector. \\
\\
\underline{K$\alpha$ emission spectra}\\
The presence of nitrogen-like ions and therefore the resonant absorption transitions in the Cu targets (\textit{case 2}) and their absence in the case of Si-Cu compound targets (\textit{case 1}) was independently verified by K$\alpha$ emission spectra. Those were measured by a spectrometer employing a 2x4 cm large HOPG crystal with a mosaicity of m=0.8° observing the rear side of the target. This instrument provides a spectral resolution better then $10\unit{eV}$, the data are spatially and temporally integrated. Fig.~\ref{fig:Kalpha} presents the measured spectra for the two different targets highlighting the K$\alpha$ emission line of N-like Cu at $8165\unit{eV}$.
\begin{figure}
    \centering
    \includegraphics[width=0.55\linewidth]{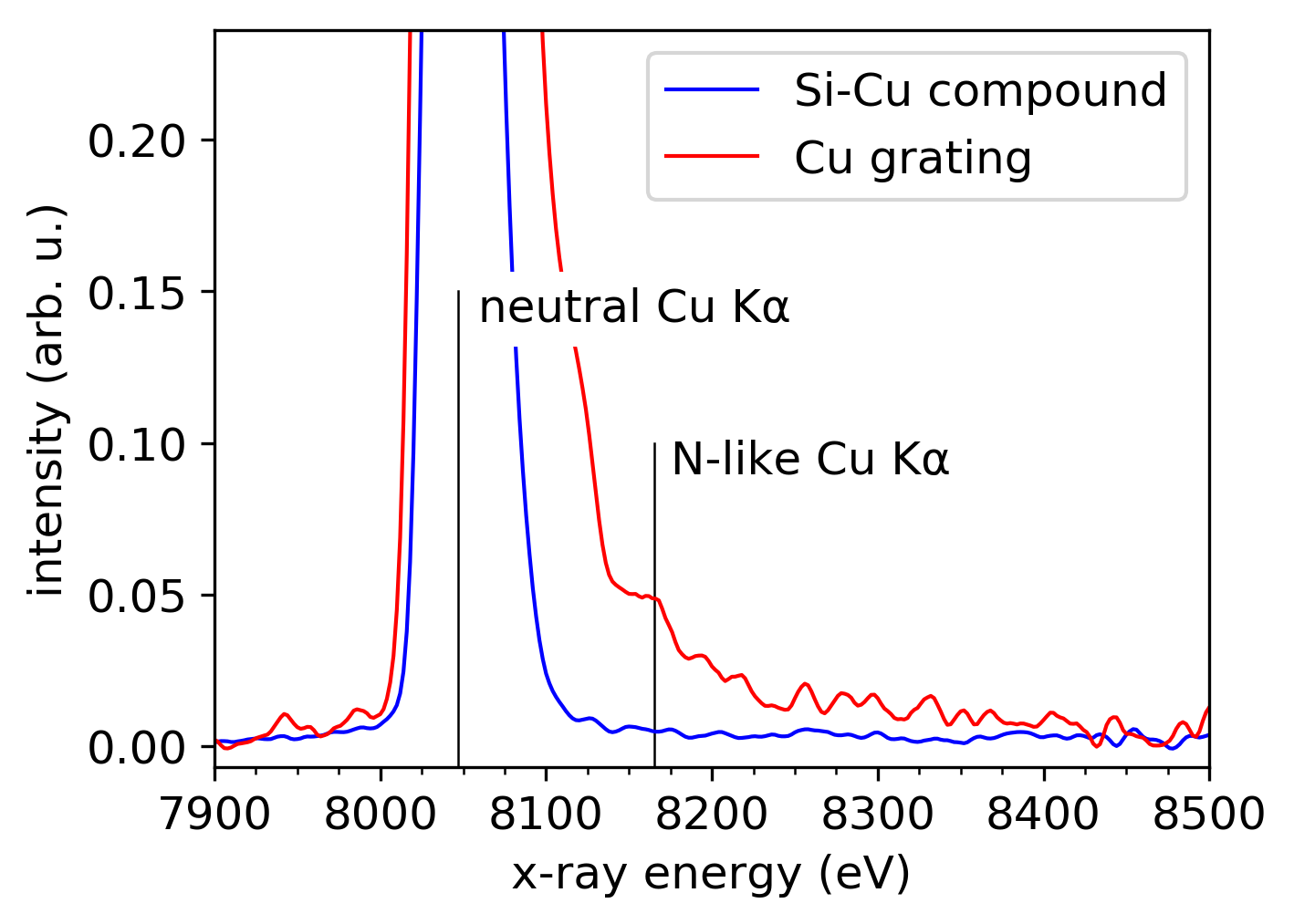}
    \caption{The measured K$\alpha$ emission spectra for two different target compositions. In the case of a pure Cu target, the emission at the probe photon  energy of $8165\unit{eV}$ is clearly visible showing the presence of Ni-like ions. In contrary, in the Si-Cu compound target the ionization of Cu is far not that high as most of the laser energy is absorbed in the Si layer.}
    \label{fig:Kalpha}
\end{figure} \\
\\
\underline{PIC simulations} \\
We performed PIC simulations using the collisional PIC code SMILEI\cite{Derouillat2018}. 
This code treats binary collisions within a cell between charged particles and includes ionization via field ionization and collisional direct impact ionization~\cite{Mishra2013}. 
It does not include recombination, hence the simulation gets unphysical when the plasma is first heated and then is cooled down too much, which is the case in case 2 after a few hundred fs. 
The spatial and temporal resolution for the simulations was set to $\Delta x=\Delta y = c\Delta t/\sqrt{2}=\lambda_{laser}/\cdot 400$ for case 1 and $\lambda_{laser}/\cdot 800$ for case 2. The simulation box was set to $3.6\unit{\mum}$ in x-direction (laser direction) for case 1, $1.6\unit{\mum}$ for case 2, and $200\unit{nm}$ in transverse y-direction. 
The laser polarization was set in y-direction, the transverse envelope was a planar. We took the finite waist into account by running the simulation several times with different laser intensity and averaging the respective contribution to the asymmetry. The other laser parameters are set to experimental values shown in Fig~1b in the main text. 
We placed $20$ ions per cell, Copper being preionized to charge state 1+, and added electrons to start with a neutral plasma.
All other geometric and laser parameters were set according to Fig.~1 in the main text. \\
\\
\underline{Hydrodynamic simulations}\\
The hydrodynamic simulations in \textit{case 1} were performed in PALE2 code\cite{Liska2011}. It is a two-dimensional Arbitrary Lagrangian Eulerian code for laser plasma simulations, but only the Lagrangian steps (i.e., the computational mesh following the flow of the matter) were applied, in order to maintain the interface between the materials perfectly resolved. The initial condition was given by the PIC results at time 0.5 ps, which were integrated over the velocity space and averaged in the transversal direction to eliminate the sampling noise of the particle method. The only exception from the rule was the density and material profile and the corresponding mesh shape, where the analytic formulae for the buried and rear interfaces were used. The computational mesh consisted of 100 cells in the transversal and 200 cells in the longitudinal direction, where exactly half of them was dedicated to each of the material parts considering their different densities. The mesh was non-uniform with the spatial step geometrically decreasing towards the interfaces with the coefficient 0.96, in order to model their thermal motion in detail. In addition to the two-temperature hydrodynamics, the electron heat diffusion model was employed, where the Spitzer-Härm formula (adopted from\cite{Lee1984}) was corrected for the dense and low temperature plasma according to\cite{Eidmann2000a} based on the bulk solid heat conductivities. The radiation transport was found to have insignificant effects on the dynamics for the given range of temperatures, so it was omitted in the simulations for simplicity and consistency with the PIC modelling. \\
\\
\underline{Modeling of the asymmetry} \\
The asymmetry was analytically modeled for Fig.~5, 6 and~7f based on Eqn. (1) in the main text as discussed below in the Supplementary Text, section \ref{sec:CalcAsy}. 
As is shown there, the asymmetry is due to the existence of hydrogen and oxides on the target surface and the small randomness in their thickness. However, the asymmetry is not very sensitive to their degree of expression.
Their existence can be experimentally shown by the data recorded separately by a Thomson parabola spectrometer, see Supplementary. 
The exact thickness cannot be determined by this method but their approximate values can be extracted from literature\cite{SebastianMader2011,NeergaardWaltenburg1995,*Hollauer2007,Belkind2008}. We take into account the imperfect target manufacturing by assuming all geometric parameters to vary randomly by $\pm 10 \%$ (ridge height, widths, smoothnesses and positions; oxidation and contamination layer thicknesses). \\
\\
\underline{Atomic rate simulations}\\
The opacity of the trajectory shown in Fig.~7f by the orange arrow was estimated by running simulations using the toolkit SCFLY\cite{scfly} with different non-LTE temperatures at a collisional radiative steady state. 
The resulting ionization distributions were compared to that of the PIC simulation at each time step and the SCFLY simulation with the best agreement were chosen. In a second step, those temperatures were used for the opacity calculation using SCFLY-spec in order to obtain an approximate temporal evolution of the expected plasma opacity. \\
\\
\underline{Calculation of the experimental average asymmetry}\\
The raw data PIXIS signal was summed within a square of 17 pixels width for each individual scatter peak and then corrected by the corresponding mean of two background values determined at both sides of the signal axis. The result was normalized to photon numbers $N_{ph}$ by a conversion factor of $1/164$ photons per ADU which was calibrated offline (as in \cite{Kluge2018}). The total uncertainty is given by the Poisson error $\sqrt{N_{ph}}$, the statistical background error given by the standard deviation $\sigma_{Bg}$ of 12 background values allocated at both sides of the peaks) and the dark field error $\sigma_{DF}$ (measured analogously). Applying Gaussian error propagation the individual peak error is then given by $\Delta N_{ph} = \sqrt{N_{ph}(q)+(\sigma_{Bg}-\sigma_{DF})^2+(\sigma_{DF})^2}$. For each pair of scattering peaks situated at $+q_i$ and $-q_i$ with photon numbers $N_{ph,i}^+$ and $N_{ph,i}^-$, respectively, the individual asymmetry was calculated according to $\eta_i=\left(N_{ph,i}^+-N_{ph,i}^+\right)/\left(N_{ph,i}^++N_{ph,i}^-\right)$. This peak-wise asymmetry was then averaged over all $q_i$ in the $q$-range that is experimentally accessible, which helps to reduce the error margin in the experiment compared to the single measurement $\eta_i$. The error of the average asymmetry is then again derived by applying Gaussian error propagation. \\
In \textit{case 1} we averaged the asymmetry $\overline{\left|\eta_0\right|}$ over the experimentally accessible $q$-range from $q = 0.05\mathrm{nm}$ to $0.15/\mathrm{nm}$, preshot asymmetries are around $0.13\pm 0.5$. 
The accessible $q$-range in \textit{case 2} was larger due to the use of absorbers, $q \approx 0.015\mathrm{nm}$ to $0.2/\mathrm{nm}$. The average preshot asymmetry here ranged from $0.03$ to $0.09$ with uncertainties in the range from 0.01 to 0.06.\\
\\
\underline{Statistical significance}\\
To determine the statistical significance of the resonant main shots on copper being more asymmetric than the preshots and the off-resonance shots being less asymmetric, we calculated the conditional probability of the respective shots being either all more asymmetric ($p_{>}$) or all less asymmetric ($p_{<}$). 
Here, $p_> = \prod_{j\epsilon M} p_>^j$ is the product of the individual probabilities for each shot $j$ of the set $M$ of resonant or off-resonant shots. For example, the probability for the asymmetry being increased on the resonant main shots is given by $p_>/(p_<+p_>)$. To calculate $p_>^j$ ($p_<^j$) we assume a Gaussian distribution of the probability density distribution for the real value being around the measured value, with the width of the Gaussian given by the respective measurement error bar, and integrate from $\eta/\eta_0=1$ to $\infty$ (from 0 to 1). The result is $p_>/(p_<+p_>)=0.999999902$ for the resonant shots with delays equal to or less than $300\unit{fs}$, which is larger than the $5\sigma$ level. 

\section{Supplementary Text}
\subsection{Modeled form factors and opacities}
\begin{figure}[h!]
\centering
      \includegraphics[width=0.6\linewidth]{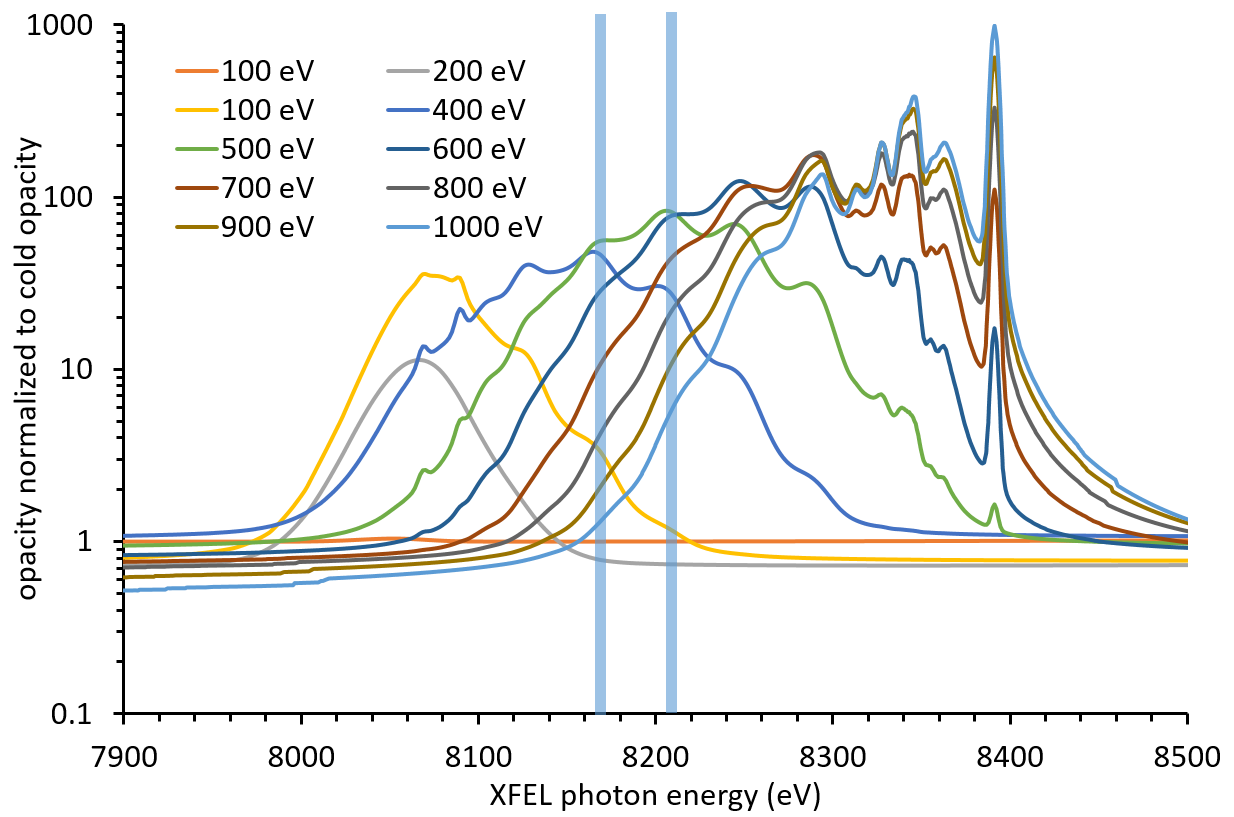}
\caption{Opacities as a function of XFEL probe energies (the energies used in the experiment are highlighted by blue bars) for different plasma temperatures, using FLYCHK\cite{flychk}.}
\label{fig:opacitiesT}
\end{figure}

\begin{figure}[h!]
\centering
      \includegraphics[width=0.5\linewidth]{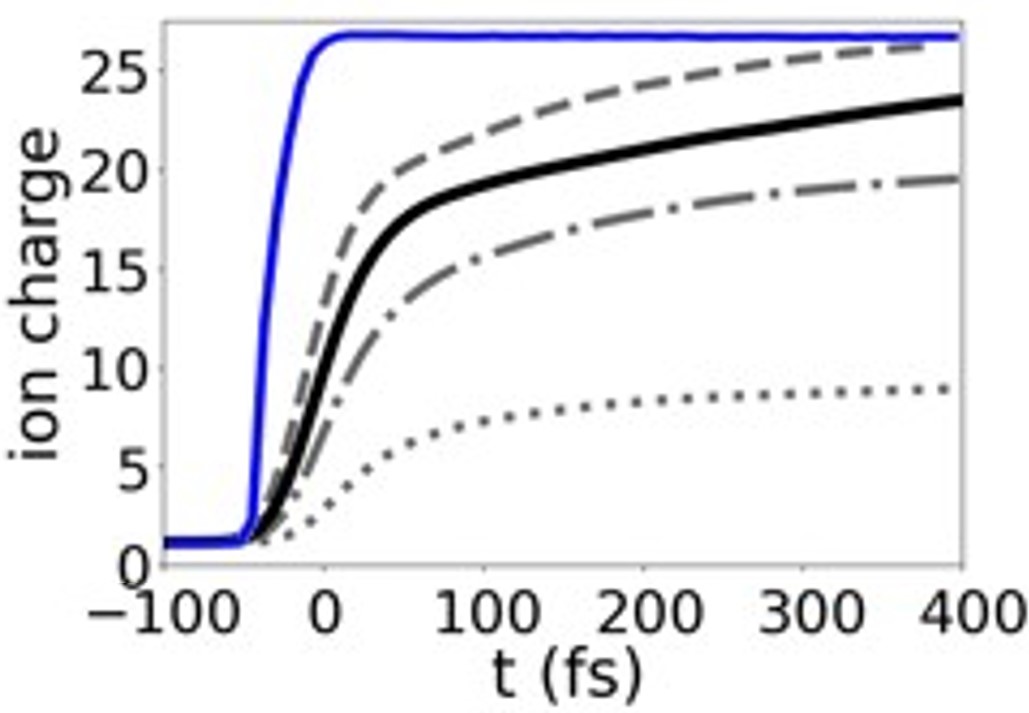}
\caption{Average Cu charge states from the PIC simulations (black) and from corresponding SCFLY runs (blue) where the time history for the electron and ion temperature was taken from the respective PIC simulation. }
\label{fig:chargeStates}
\end{figure}

Since the PIC simulations can not calculate the plasma opacity selfconsistently, we have to resort to approximate the opacities using a separate software SCFLY\cite{scfly}. 
Opacities were calculated using SCFLY\cite{scfly} for a range of plasma temperatures in LTE, see~\ref{fig:opacitiesT}. 
However, since SCFLY like any available atomic rate solver can not take into account the multidimensional and kinetic non-thermal and transient nature of the plasma, we cannot simply identify the NLTE temperature used in SCFLY with the average electron energy in the PIC simulation. 
This becomes obvious when comparing the average Copper charge states from SCFLY and PIC.  
The SCFLY-runs, with the temporal evolution of the electron and ion temperatures extracted from the PIC simulations, considerably overestimate the charge state especially in the first few hundred femtoseconds where we did our measurements, cf. Fig.~\ref{fig:chargeStates}. 
Therefore, in order to approximate the opacities we took the approach to rather match the average charge states for each PIC timestep from a series of SCFLY runs with different NLTE temperature. 
From the matching SCFLY run we then extracted the respective opacity and finally obtain the evolution plotted in the bottom graph of Fig.~7e in the main text. 



\subsection{Shot raw data}
\subsubsection{Si-Cu compound gratings, case 1}
\begin{figure}
\centering
      \includegraphics[width=0.9\linewidth]{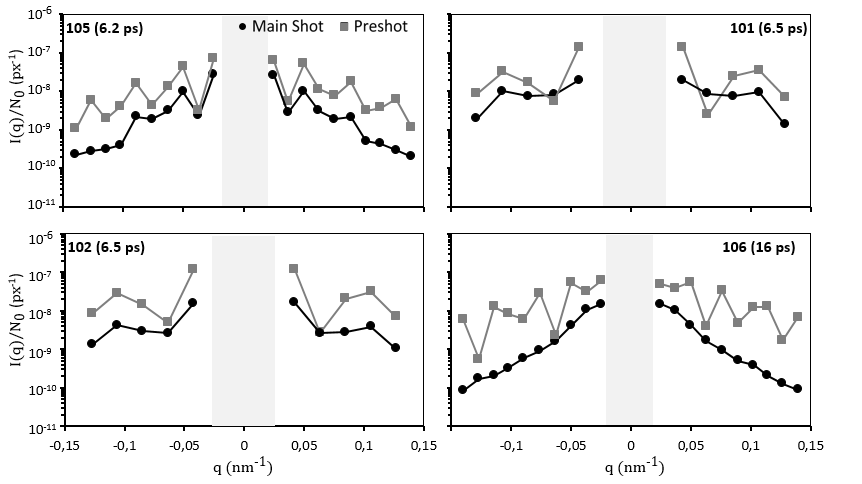}
\caption{Scattering amplitudes in the first maxima, normalized to the number of photons in the respective XFEL pulse. Gray: asymmetric XFEL-only preshot, black: symmetric main shot with HI laser + XFEL. Lines are guides to the eye. The gray shaded area could not be evaluated due to over-saturation. Note, that due to the semi-logarithmic scale asymmetries visually appear much smaller than they are. Significant reduction of intensity at large q, connected to an increase of the grating smoothness, can be seen only after 16 ps.}
\label{fig:maxima_low}
\end{figure}

\begin{figure}
\centering
      \includegraphics[width=0.9\linewidth]{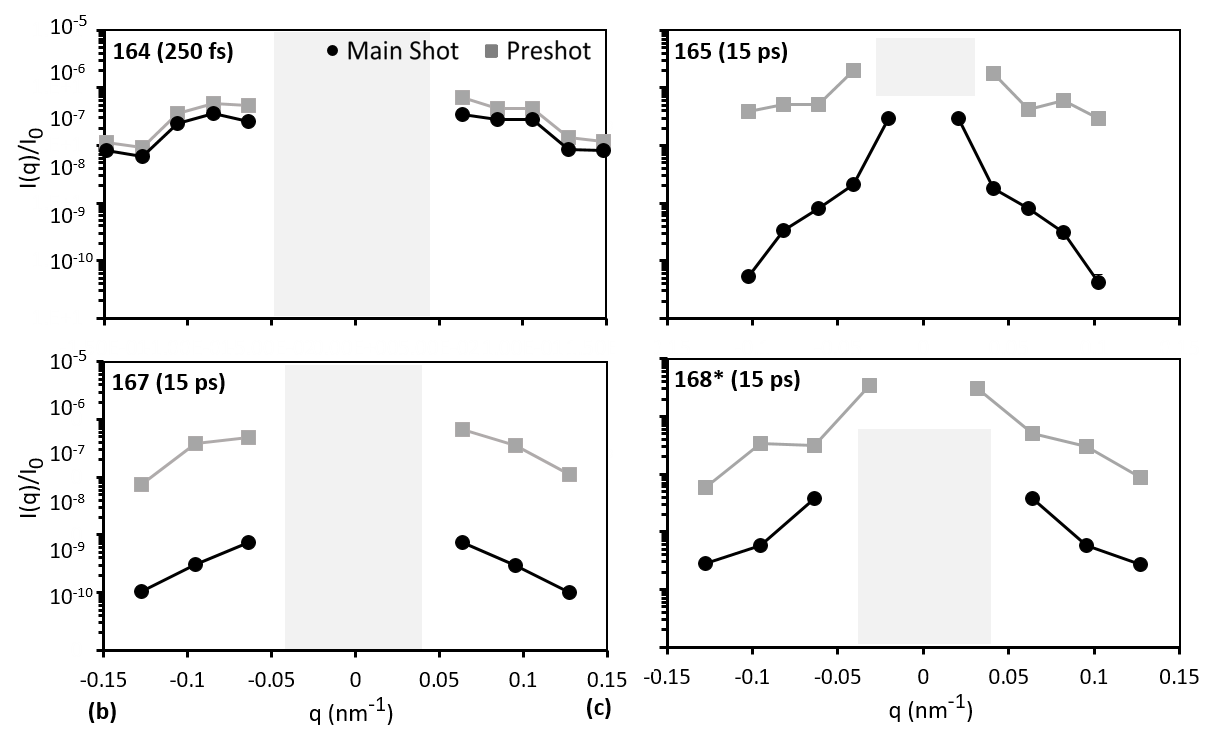}
\caption{Scattering amplitudes in the first maxima, normalized to the number of photons in the respective XFEL pulse. Gray: asymmetric XFEL-only preshot, black: symmetric main shot with HI laser + XFEL. Lines are guides to the eye. The gray shaded area could not be evaluated due to over-saturation. Note, that due to the semi-logarithmic scale asymmetries visually appear much smaller than they are. (*shot 168: $E_0=8000\unit{eV}$).}
\label{fig:maxima}
\end{figure}

The integrated peak heights for \textit{case 1} are shown in Fig.~\ref{fig:maxima_low} and~\ref{fig:maxima} for $0.4\unit{J}$ and $1\unit{J}$, respectively. 

\newpage

\newpage

\subsubsection{Cu gratings, case 2}

\begin{figure}
\centering
      \includegraphics[width=0.85\linewidth]{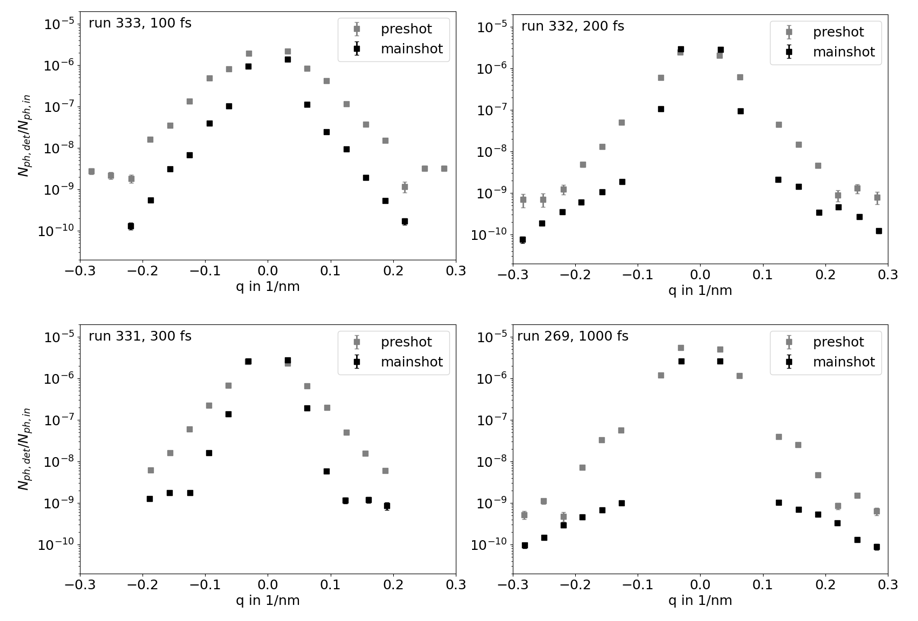}
\caption{Scattering amplitude in the first maxima, normalized to the number of photons in the respective XFEL pulse for exemplary shots on Cu targets at resonant XFEL energy.}
\label{fig:CuRes}
\end{figure}

\begin{figure}
\centering
      \includegraphics[width=0.85\linewidth]{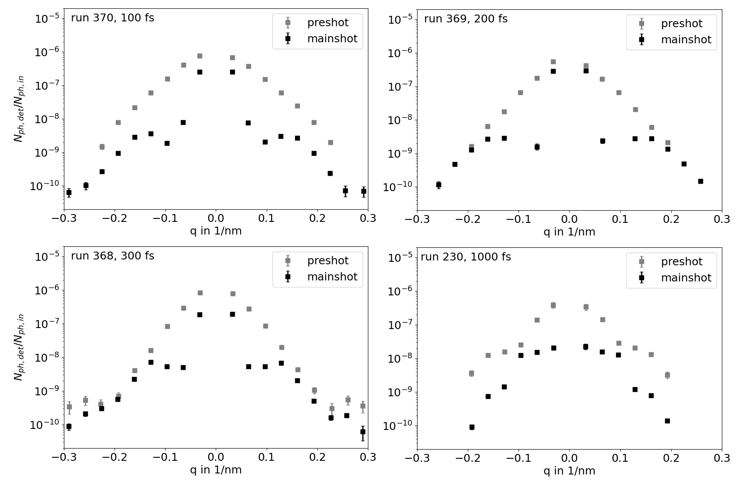}
\caption{Scattering amplitude in the first maxima, normalized to the number of photons in the respective XFEL pulse for exemplary shots on Cu targets at off-resonant XFEL energy.}
\label{fig:CuOffRes}
\end{figure}

\begin{figure}
\centering
      \includegraphics[width=0.6\linewidth]{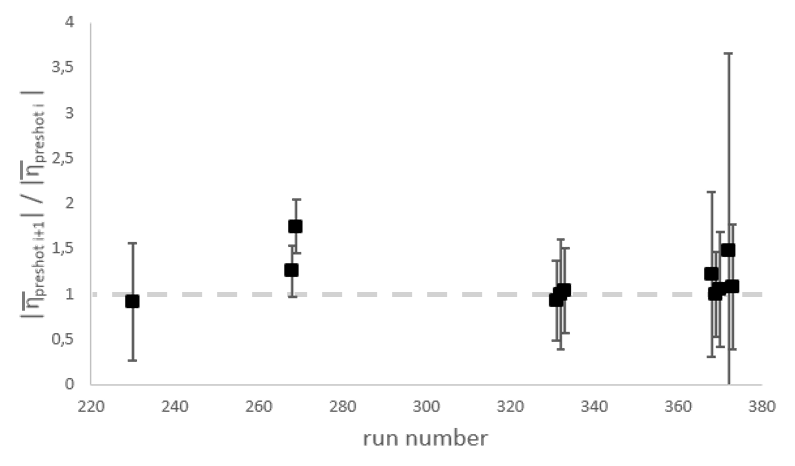}
\caption{Ratio of the average preshot asymmetry between two consecutive preshots on the same Cu grating target indicated by their respective run number. The scattering signals of preshots on Cu grating targets yield similar average asymmetries and hence ratios close to 1.}
\label{fig:Case2PreshotRatio}
\end{figure}

Fig.~ \ref{fig:CuRes} and Fig.~\ref{fig:CuOffRes} show exemplary shots from the data set of shots on Cu targets at resonant and off-resonant XFEL energy. Consecutive preshots on the same Cu grating exhibit similar average asymmetry values in the scattering patterns. Hence the ratio of the average asymmetry in the scattering signal of two consecutive preshots is close to 1 as can be seen in Fig~\ref{fig:Case2PreshotRatio}. This means that differences in the asymmetry between preshots and main shots cannot be explained by a random variation of the XFEL phase, as this would influence also the preshots. 
The fact that the changes of the asymmetry we describe in the main text for the main shots are not occuring in the preshots hence means that those changes are in fact correlated to the HI laser generated plasma dynamics rather than the XFEL-only properties.  

\newpage

\newpage

\newpage

\begin{figure}
\centering
      \includegraphics[width=0.85\linewidth]{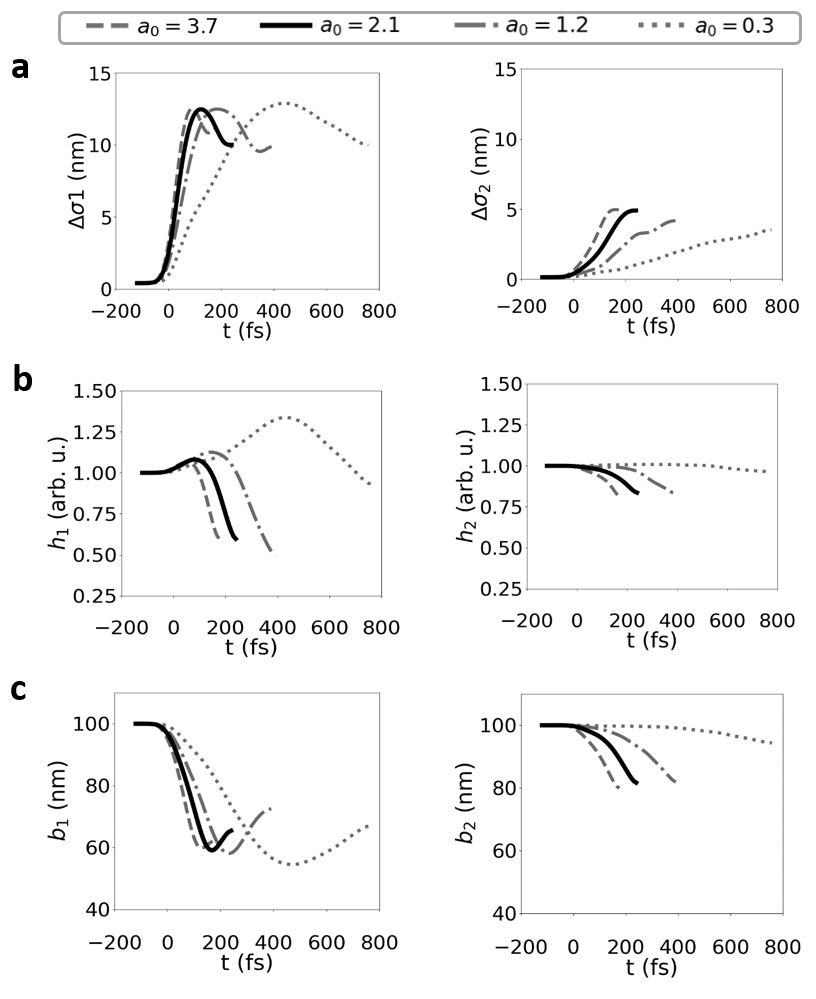}
\caption{Dynamics of the geometrical parameters of the front (1) and rear (2) grating obtained from the Smilei simulations for different values of the relativistic laser amplitude $a_0$. The fit parameters were obtained by fitting the integrated total electron density in the front and rear half of the target with error functions ($z\propto h_{1,2} \cdot  \mathrm{erf}\left[\left(y-y_0\right)/\left(\sqrt{2}\sigma_{1/2}\right)\right]$). \textbf{a} relative smoothing of the edges $\Delta \sigma_{1,2}=\sigma_{1,2}(t)-\sigma_{1,2}(t_0)$, \textbf{b} grating amplitudes $h_{1,2}$ and \textbf{c} width of the grating ridges $b_{1,2}$ }
\label{fig:gratingDynamics}
\end{figure}

Fig.~\ref{fig:gratingDynamics} shows the development of the geometrical parameters of the front and rear side grating obtained from the PIC simulations. The development of the parameters of the grating edge smoothness $\sigma_{1,2}$ and the amplitudes $h_{1,2}$ were used as input for our model to calculate the asymmetry trajectory in Fig. 7f in the main text.


\newpage

\subsection{Existence of surface oxides}
\begin{figure}
\centering
      \includegraphics[width=0.85\linewidth]{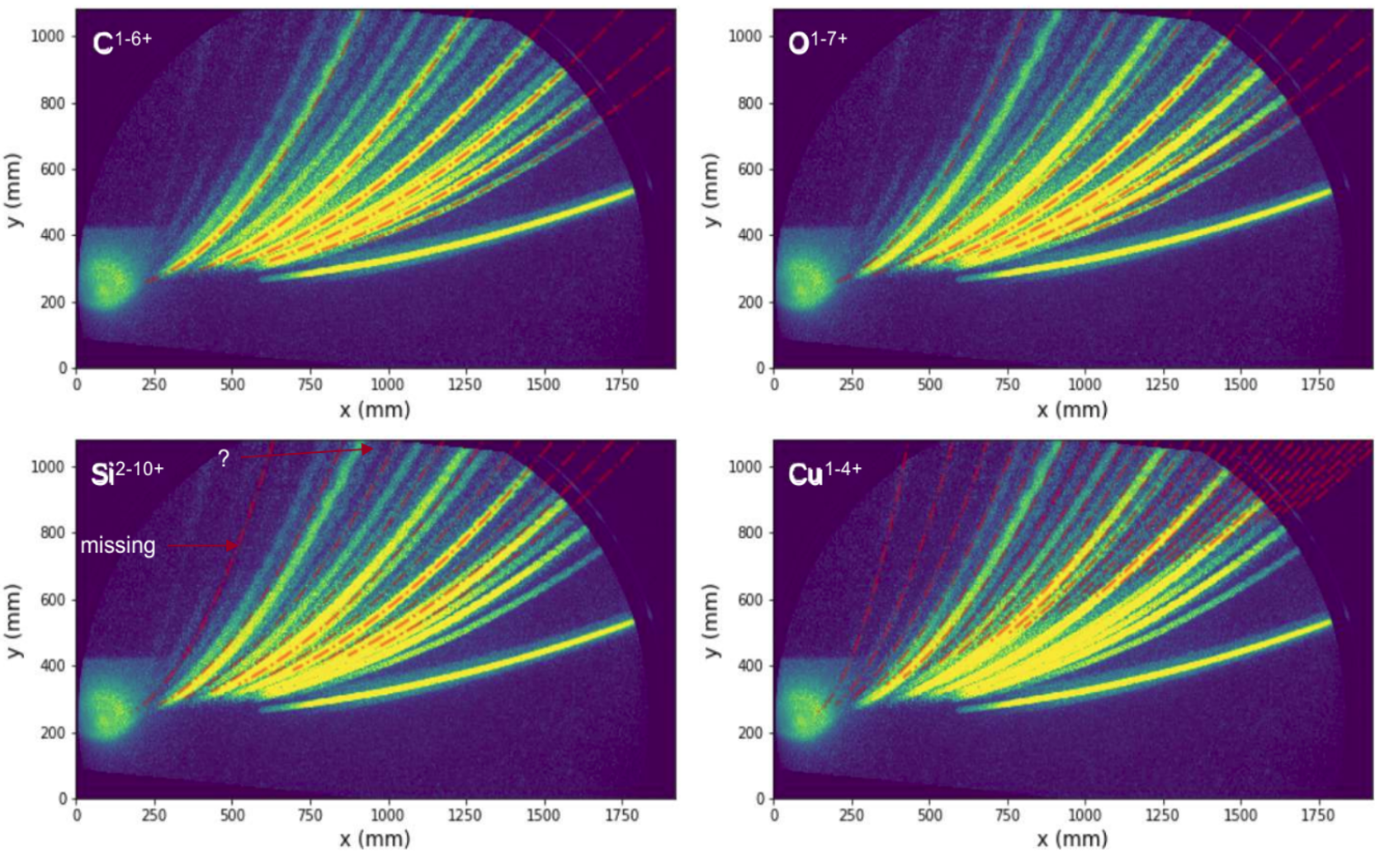}
\caption{Thomson parabola traces and the respective trace identification as C and O ions. Also, there are traces of $H^+$, $Cu^{1+}-Cu^{4+}$ as well as remnants of the Si carrier membrane visible.}
\label{fig:CuOxide}
\end{figure}

The existence of oxides on the target surface can be shown by the data recorded by a Thomson parabola. 
It was positioned behind the target along the target normal direction with a distance of approximately $93\unit{cm}$ between the ion source and the $500\unit{\mum}$-pinhole. 
The ion traces for an example shot on a copper grating target are shown in Fig.~\ref{fig:CuOxide}.
They show the existence of oxygen ions in the accelerated ion cloud, which means that oxygen was present on the surfaces either in a layer of hydro-carbon compounds and/or CuO on the surface. 
There is also a strong hydrogen line visible, giving evidence to a hydro-carbon compound layer on the target surface.

\subsection{\label{sec:CalcAsy}Calculation of the asymmetry}
We start with a multilayer target with $N$ layers with arbitrary interface shapes $z_i\left(\vec{r}\right)$ where $i$ is the index of the layer interface and $\vec{r}=\left(x,y\right)$ is a vector within the plane perpendicular to the X-ray direction $z$. 
By this definition, layer $i$ is located between $z_{i-1}\left(\vec{r}\right)$ and $z_i\left(\vec{r}\right)$, the front surface of the target is described by $z_{0}\left(\vec{r}\right)$. 
The conventional definition of the replication factor of interface $i$ with respect to the interface $j$ is defined by the ratio of the power spectrum distributions (PSD) of layer $i$ and $j$, and reads
\begin{equation}
    R_{i,j}\left(\vec{q}\right) = \frac{\left| \tilde z_i\left(\vec{q}\right) \right|^2}{\left| \tilde z_{j}\left(\vec{q}\right) \right|^2}. 
\end{equation}
Throughout the paper we use the tilde symbol to indicate the Fourier transform, $\tilde z\left(\vec{q}\right) = \mathcal{F}_{\vec{r}}\left[z\left(\vec{r}\right)\right]\left(\vec{q}\right)$. 
The vector $\vec{q}=\left(q_x,q_y\right)$ denotes the scattering vector $\vec{q}=\vec{k'}-\vec{k_0}$ where $\vec{k_0}$ is the wave vector of the incoming X-ray beam and $\vec{k'}$ is that of the scattered beam, assuming the scattering angle is small and the scattering process is sufficiently elastic, so that $q_z\approx 0$. 
The replication factor as defined above is 1 for perfect replication, i.e. $\tilde z_i\left(\vec{q}\right) = \tilde z_{j}\left(\vec{q}\right)$ and can take on values between 0 (for frequencies contained only in $\tilde z_{j}$) and infinity (for frequencies contained only in $\tilde z_i$). 
For simplicity, in the following we use the term \textit{replication factor} for the complex-valued ratio
\begin{equation}
    \chi_{i,j}\left(\vec{q}\right) = \frac{\tilde z_{i}\left(\vec{q}\right)}{\tilde z_{j}\left(\vec{q}\right) }.
\end{equation}
Its difference between two adjacent interfaces reads
\begin{equation}
    \Delta\chi_{i,j}\left(\vec{q}\right)\equiv\chi_{i,j}\left(\vec{q}\right)-\chi_{i-1,j}\left(\vec{q}\right).
\end{equation}
With the definition $\Delta\chi_{i}\equiv\Delta\chi_{i,0}$ the difference between the Fourier components of the interface $i$ and $i-1$ is simply given by $\tilde{z}_i-\tilde{z}_{i-1}=\Delta\chi_i\left(\vec{q}\right) \tilde z_{0}\left(\vec{q}\right)$. 

\begin{figure*}
    \centering
    \includegraphics[width=\linewidth]{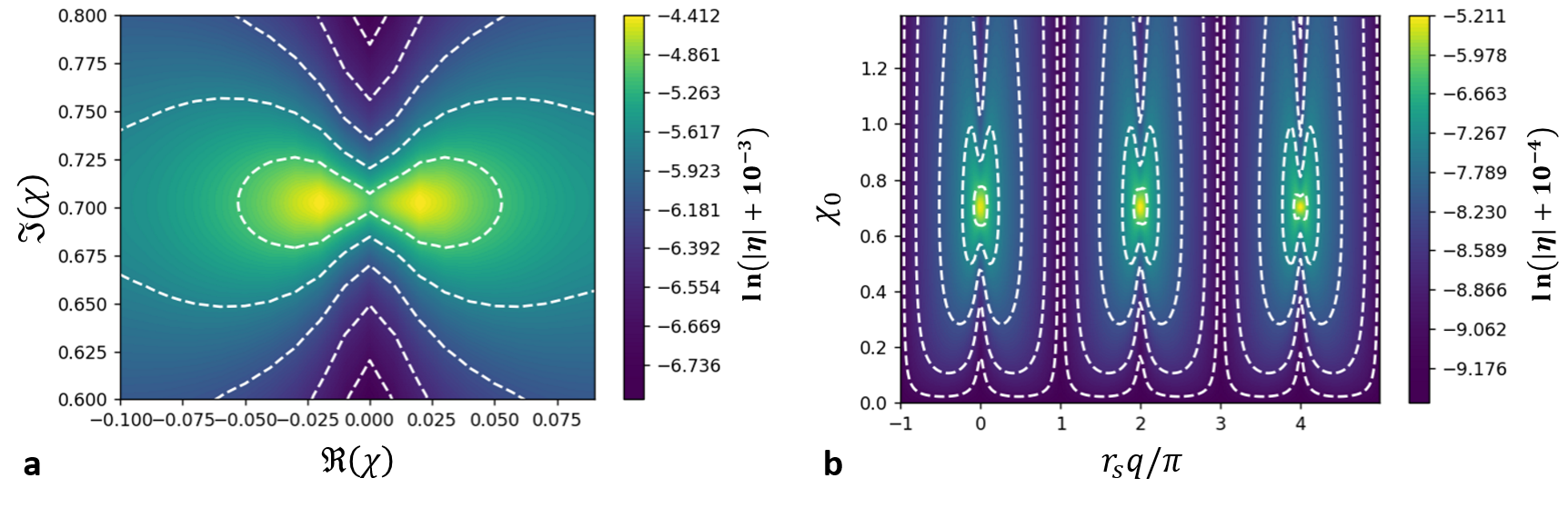}
    \caption{Absolute value of the asymmetry from Eqn.~\eqref{eqn:eta2layer} (a) and Eqn.~\eqref{eqn:eta2layerSigma} for a Si-Cu compound with flat Si front surface. }
    \label{fig:etaSiCu}
\end{figure*}
The intensity $I\left(\vec{q}\right)$ of the scattered light in Born approximation is proportional to the Fourier transform of the absolute square of the scattering length areal density $\zeta\left(\vec{r}\right)$ multiplied with the complex-valued X-ray field amplitude $a_0\left(\vec{r}\right)$, $I\left(\vec{q}\right)\propto \left|\mathcal{F}\left[a_0\left(\vec{r}\right)\zeta\left(\vec{r}\right)\right]\left(\vec{q}\right)\right|^2$ with
\begin{equation}
    \zeta\left(\vec{r}\right)=\sum_{i=1}^N \rho_i\left(\vec{r}\right)\left[z_i\left(\vec{r}\right) - z_{i-1}\left(\vec{r}\right)\right]
    \label{eqn:zeta}
\end{equation}
and
\begin{equation}
    \rho_i\left(\vec{r}\right)=n_i\left(\vec{r}\right)\left(f_i^0+f'_i+\boldmath{i}f''_i\right). 
\end{equation}
Here, $n_i$ denotes the atom or ion density in layer $i$ and $f_i^0$,  $f'_i$,  $f''_i$ are the ion form factor and optical corrections. 
In small angle approximation it is $f_i^0\approx Z$, where $Z$ is the atomic number of the atom or ion. 
Note, that this equation is valid without the loss of generality, as layers consisting of mixtures of atoms or ions with different atomic numbers and/or different optical corrections could be easily represented by the respective average values. 
In the following we assume that the materials of each layer are homogeneous, i.e. the density is not a function of $\vec{r}$, $\rho_i\left(\vec{r}\right)=\rho_i$. 
Then the Fourier transform and its conjugate of Eqn.~\eqref{eqn:zeta} read
\begin{eqnarray}
    \tilde  \zeta\left(\vec{q}\right)&=&\sum_{i=1}^N \rho_i\left[\tilde z_i\left(\vec{q}\right) - \tilde z_{i-1}\left(\vec{q}\right)\right]\\
    \tilde  \zeta\left(-\vec{q}\right)&=&\sum_{i=1}^N \rho_i\left[\tilde z^*_i\left(\vec{q}\right) - \tilde z^*_{i-1}\left(\vec{q}\right)\right]. 
\end{eqnarray}
The latter can be easily seen from splitting $z_k$ into their even and odd components $z_{k,e}$, $z_{k,o}$, respectively: The Fourier transform of an even real-valued function is again even and real-valued, while that of an odd real-valued function is odd and imaginary, hence
\begin{eqnarray}
    \nonumber
    \tilde z_i\left(\vec{q}\right)&=&\tilde z_{i,e}\left(\vec{q}\right)+\tilde z_{i,o}\left(\vec{q}\right)\\
    \nonumber
    &=&\tilde \Re{\left[z_{i,e}\left(\vec{q}\right)\right]}+i\Im\left[\tilde z_{i,o}\left(\vec{q}\right)\right]\\
    \tilde z_i\left(\vec{-q}\right)&=&\Re{\left[\tilde z_{i,e}\left(\vec{q}\right)\right]}-i\Im\left[\tilde z_{i,o}\left(\vec{q}\right)\right]=\tilde z^*_i\left(\vec{q}\right)
\end{eqnarray}

We will now analyze how the normalized asymmetry 
\begin{equation}
    \eta\left(\vec{q}\right)=\frac{I\left(\vec{q}\right)-I\left(-\vec{q}\right)}{I\left(\vec{q}\right)+I\left(\vec{-q}\right)}
\end{equation}
is related to the replication factors. 
The above definitions allow us to give this equation explicitly in terms of the interface layer geometries in a very compact expression if $a_0=const.$: 
\begin{equation}
    \eta\left(\vec{q}\right) = 
    \frac{\left| \sum_{i=1}^N \rho_i \, \Delta\chi_i\left(\vec{q}\right) \right|^2 - \left| \sum_{i=1}^N \rho_i \, \Delta\chi^*_i\left(\vec{q}\right) \right|^2}
         {\left| \sum_{i=1}^N \rho_i \, \Delta\chi_i\left(\vec{q}\right) \right|^2 + \left| \sum_{i=1}^N \rho_i \, \Delta\chi^*_i\left(\vec{q}\right) \right|^2}
    \label{eqn:eta_start}
\end{equation}
which can be rewritten as 
\begin{equation}
    \eta\left(\vec{q}\right)=\frac{\sum_{i=1}^N \sum_{j=1,j\neq i}^N a_{ij} \Im{\left(\Delta\chi_i^*\left(\vec{q}\right) \, \Delta\chi_j\left(\vec{q}\right)\right)}}{\sum_{i=1}^N\sum_{j=1}^N b_{ij}\Re{\left(\Delta\chi_i^*\left(\vec{q}\right) \, \Delta\chi_j\left(\vec{q}\right)\right)}}.
    \label{eqn:etaFinal}
\end{equation}
This equation connects the asymmetry with the replication factors and the material-dependent factors 
\begin{eqnarray}
    a_{ij}&=&2\left(f_j^0+f_j'\right)f_i'' n_i n_j\\
    b_{ij}&=&\left[\left(f_j^0+f_j'\right)\left(f_i^0+f_i'\right)+f_i''f_j''\right] n_i n_j. 
\end{eqnarray}
Most notably, this relation is independent of the specific geometry of the layer interfaces, i.e. all $\tilde z_i$ have completely cancelled out. 
Only the relative differences, described by the replication factors, are important. \\
Secondly, the replication factors need to be complex-valued, i.e. the imaginary part must not vanish completely for all replication factors in order for the asymmetry to be non-zero. 
A simple consequence is that the asymmetry is constant when the target explodes symmetrically. When it expands longitudinally, the densities reduce, but since $a_{ij}$ and $b_{ij}$ are both quadratic in $n$, this change cancels out. 
When the interfaces expand laterally, i.e. $\vec{r}'=\vec{c_i}\vec{r}$ (here the prime denotes the positions after expansion), then the replication is only real-valued, $\Delta\chi_i\left(\vec{q}\right)=\tilde z_i\left(q/c_i\right)/c_i\tilde z_i\left(q\right)$ and therefore even. \\
Thirdly, at least two layers are needed to generate an asymmetry, since otherwise the sum over $j\neq i$ in the numerator is not defined. 
This is physically reasonable since the asymmetry is generated by a difference between the spatial areal  distribution of the real part of the form factors and optical corrections on the one hand, and the imaginary part of the optical corrections on the other hand. 
Since by definition there is no difference between them in one layer (up to constant factor), at least two layers are needed. 
Additionally, in order to generate asymmetry, for each pair of two layers it is required that the ratios between the real and imaginary parts are different, i.e. $a_{ij}\neq a_{ji}$. \\

\subsubsection{Two layer targets}
The most simple case in which the asymmetry is not vanishing is that of a two-layer compound target where one of the target surfaces is perfectly flat, i.e. we choose $z_0\left(\vec{r}\right)=const.$ 
This setup allows us to write the asymmetry in the simple form
\begin{equation}
    \eta\left(\vec{q}\right)=\frac{c_1 \Im\left(    \chi\left(\vec{q}\right)\right)}{\Re\left[c_2 \left|\chi\left(\vec{q}\right)\right|^2+c_3 \chi\left(\vec{q}\right) + c_4\right]}
    \label{eqn:eta2layer}
\end{equation}
with the material constants
\begin{eqnarray*}
    c_1&=&2n_1 n_2 \left[(f_1^0+f'_1)f''_2-(f_2^0+f'_2)f''_1\right]\\
    c_2&=&n_2^2\left[\left(f_2^0+f'_2\right)^2+\left(f''_2\right)^2\right]\\
    c_3&=&2 n_1 n_2 \left[\left(f_1^0+f'_1\right)\left(f_2^0+f'_2\right)+f''_1f''_2\right]-c_2\\
    c_4&=&n_1^2\left[\left(f_1^0+f'_1\right)^2+f''_1\right]-2\left(c_2+c_3\right)
\end{eqnarray*}
and $\chi\left(\vec{q}\right)\equiv \chi_{2,1}\left(\vec{q}\right)$. 
This equation connects the experimentally accessible asymmetry function $\eta\left(\vec{q}\right)$ with the material properties (form factors and optical corrections in the factors $c_i$) and the replication factor of the rear surface with respect to the buried interface $\chi\left(\vec{q}\right)$. 

In order to give a quantitative example we plotted the calculated asymmetry of a silicon-copper compound target in Fig.~\ref{fig:etaSiCu}a, assuming a flat silicon front surface and a structured Si-Cu interface and rear surface. 
Such a target can be produced e.g. by structuring the rear side of a Si wafer and covering it with a copper layer. 
For reference, the constants in this case are $c_1=3.78\times 10^{-3}$, $c_2=14.9$, $c_3=-6.0$, $c_4=-13.6$. 
We find that the asymmetry is very close to zero everywhere with the exception of distinct extrema at $\Im\left(\chi\right)\approx \pm 0.7$ and $\Re\left(\chi\right)\approx \pm 0.025$. 

\subsubsection{Example replication factors}
\underline{Shift and convolution with an even function}

We will now discuss a few specific examples for the replication factor. 
We first assume that the rear surface copies the buried interface up to a convolution with an even function. 
This requires that in Fourier space this function $\chi_0\left(\vec{q}\right)$ is even and real-valued. 
In order to break the symmetry we introduce a shift of the rear surface with respect to the buried interface by $\vec{r}_s$ which could originate e.g. due to a small tilt of the target. 
A shift in real space is expressed in reciprocal space by a multiplication with
\begin{equation}
    \chi_{s}\left(\vec{q}\right) = e^{-i\vec{r}_s \vec{q} }.
    \label{eqn:shift}
\end{equation}
The replication factor of the rear surface is then given by $\chi\left(\vec{q}\right)=\chi_0\left(\vec{q}\right)\chi_s\left(\vec{q}\right)$ and Eqn.~\eqref{eqn:eta2layer} can be further simplified to 
\begin{equation}
    \eta\left(\vec{q}\right)=\frac{c_1 \chi_0\left(\vec{q}\right)\sin{\left(\vec{r}_s\vec{q}\right)}}{c_2 \chi^2_0\left(\vec{q}\right)+c_3 \chi_0\left(\vec{q}\right)\cos{\left(\vec{r}_s\vec{q}\right)} + c_4}. 
    \label{eqn:eta2layerSigma}
\end{equation}
We plot this relation in Fig.~\ref{fig:etaSiCu}b. 
It can be readily solved for $\chi\left(\vec{q}\right)$, which demonstrates that the asymmetry of the scattering pattern directly maps the replication factor in $\vec{q}$-space. 
\\
\underline{Shift and smoothing}\\
We now assume the specific function
\begin{equation}
    \chi_0\left(\vec{q}\right) = \hat \chi e^{ - \sigma^2 q^2 }, 
\end{equation}
i.e. the rear surface copies the buried interface up to a proportionality factor $\hat \chi$ and a smoothing/sharpening by $\sigma$. 
Real-valued $\sigma$ correspond to smoothing at the rear, while imaginary values correspond to sharpening. 

\begin{figure*}
    \centering
    \includegraphics[width=\linewidth]{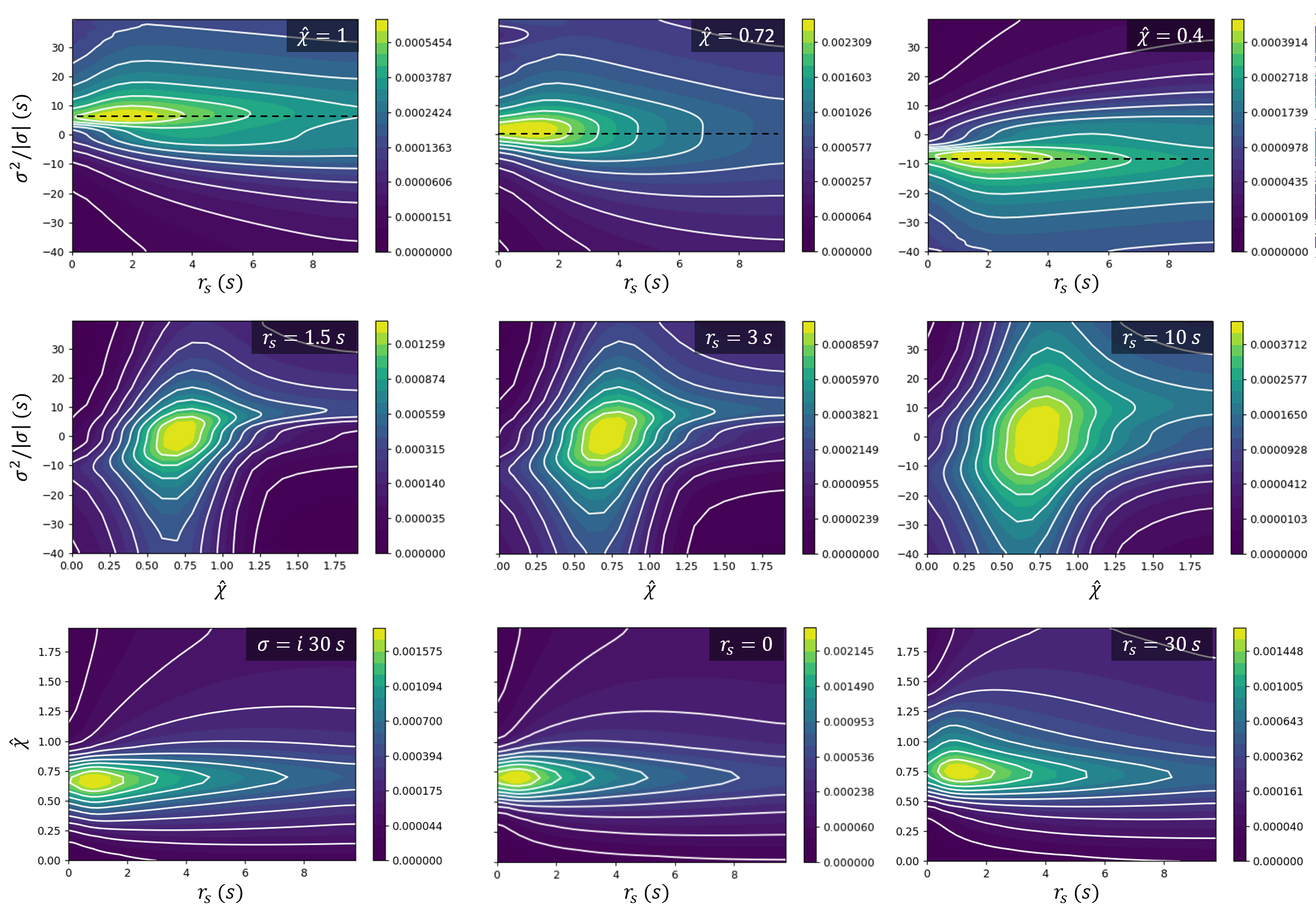}
    \caption{Average asymmetry in the $q$-range from $0-0.3 s^{-1}$ (as for all figures below) for a two layer Si-Cu compound with flat Si front surface and $\chi\left(\vec{q}\right)=\hat \chi \exp\left( - \sigma^2 q^2 - i\vec{r}_s \vec{q}\right)$, scanning the three parameters $\sigma$, $\hat\chi$ and $r_s$.}
    \label{fig:2layersTilt}
\end{figure*}

The asymmetry as a function of the three independent parameters determining the replication factor is plotted in Fig.~\ref{fig:2layersTilt}. 
Here and in the following we average the absolute value of $\eta(q)$ over the $q$-range from $0-0.3 s^{-1}$, where $s$ is the unit spatial dimension (e.g. $1\unit{nm}$). 
From the Figure we can see that the maximum asymmetry occurs when $\sigma=0$ and $\hat\chi\approx 0.7$ to $0.75$ and $r_s\approx 1.5\unit{s}$. 
Here and in the following $s$ is the unit length. 
The specific optimum parameters are material-dependent, but vary only little. 
For example, increasing the imaginary part or density of either material by an order of magnitude only changes the optimum geometric parameters by the order of $10\%$. 
Furthermore, the optimum value of the replication factor amplitude $\hat \chi_{opt}\approx 0.7$ is almost independent of $\vec{r}_s$ and $\sigma$. 

\begin{figure*}
    \centering
    \includegraphics[width=\linewidth]{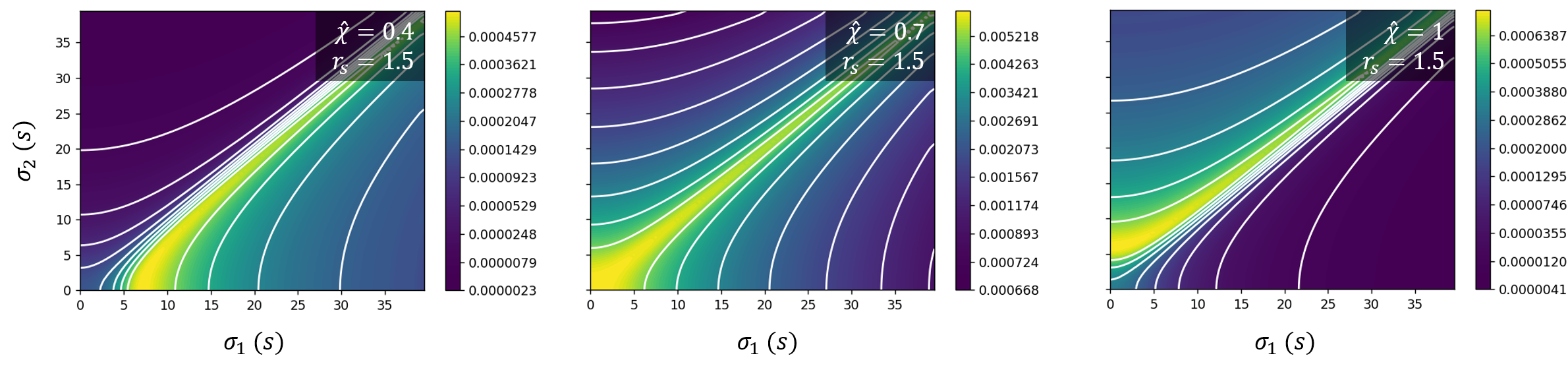}
    \caption{Average asymmetry for a two layer Si-Cu compound with flat Si front surface and $\chi\left(\vec{q}\right)=\hat \chi \exp\left( - \sigma^2 q^2 + i\vec{r}_s \vec{q}\right)$, $\sigma^2=\sigma^2_2-\sigma_1^2$ describes the relative smoothing. This exemplifies the specific case where both the buried interface (index 1) and rear surface (index 2) are described by the same function $z(\vec{r})$ and are each subject to a respective smoothing factor in Fourier space $\exp{\left(-\sigma_i q^2\right)}$. }
    \label{fig:sigmaScanTilt}
\end{figure*}
Obviously, one interesting application of the asymmetries is to experimentally restrict the possible range of the geometric parameters. 
Additionally, the change of the asymmetry upon an external drive can also be used to restrict the change of the replication factor. 
For example, if the rear layer surface initially perfectly replicates the buried interface ($\hat\chi=1$, $\sigma=0$) then from Fig.~\ref{fig:2layersTilt} one finds that an increase in the average asymmetry is equivalent to a decrease in $\chi$. 
If the rear layer does not perfectly replicate the buried interface, e.g. $\hat \chi < \hat \chi^{opt}$ or $\sigma > \sigma^{opt}$, then this relation is inverted, i.e. an increase of the average asymmetry means an increase in $\chi$. 


An important case is the study of buried layer smoothing. 
In conventional transmission SAXS the Fourier components of the buried and rear surface correlations overlap on the detector and cannot be reconstructed separately. 
The asymmetry offers a way to obtain the smoothing parameter $\sigma$ which -- assuming that upon pumping both target interfaces can only smooth out further -- can be used to extract the smoothing of the buried and rear interface separately, see Fig.~\ref{fig:sigmaScanTilt}. 
It can be seen and from the above considerations it follows that depending on $\hat\chi$ a decrease in asymmetry corresponds to a dominant smoothing of the rear surface and an increase in asymmetry corresponds to a dominant smoothing of the buried interface (a) or vice versa (c). \\
Generally, we find the following three laws:

§1: In a two-material sandwich target with a structured front surface, flat interface, and a flat rear surface that can be assumed to remain flat (e.g. for an infinitely thick target), when upon an external pump additional $q$-components arise in the transmission scattering pattern, the scattering pattern becomes asymmetric if (and only if) the buried interface changes or when the material properties become transversely modulated. 

§2: In a two-material target with a flat front surface and the replication factor of the rear surface with respect to the buried interface being separable into a real-valued function $\chi_0\left(\vec{q}\right)$ and a shift, there exists an optimum $\chi_0^{opt}$ where the asymmetry in the SAXS pattern maximizes. This optimum is independent of the shift and $\vec{q}$. The maximum asymmetries can be found close to $\vec{r}_s\vec{q}=n2\pi$. 

§3: If the rear layer is the higher Z material and the rear surface is the exact copy of the buried interface up to a shift, an increase in asymmetry is equivalent to a decrease in the replication factor and a decrease in asymmetry is equivalent to an increase in the replication factor. Further assuming both the buried interface as well as the rear surface can only \textit{increase} in smoothness, this translates into an equivalence of increasing asymmetry with increasing smoothness of the rear layer, and of a decreasing asymmetry with increasing smoothness of the buried interface. If it is not an exact copy from the beginning, this can be the other way around (i.e. when $\hat\chi<\hat\chi^{opt}$ or $\sigma>\sigma^{opt}$). 
Everything is the other way around when the real layer is the lighter Z material. 
\\
\underline{Noise}\\
So far we have only discussed the case of a perfect target. 
However, it is important to realize that small random fluctuations in the target geometry can also lead to a symmetry breaking, i.e. they do not necessarily cancel out and can take the role of the small shift $\vec{r}_s$ from the former subsection. 
However, simple fluctuations such as introducing a set of different target areas with different $\vec{r}^i_s$ or different $\chi^i_0$ in fact would tend to cancel out due to the linear and odd nature of the numerator of Eqn.~\eqref{eqn:eta2layerSigma}. 
What is needed to generate an asymmetry are non-compensated shifts. 
Since such local shifts cannot be expressed by a universal replication factor in Fourier space, independent on the specific shape of the interfaces, we have to resort to the study of example geometries. 

\begin{figure*}
    \centering
    \includegraphics[width=\linewidth]{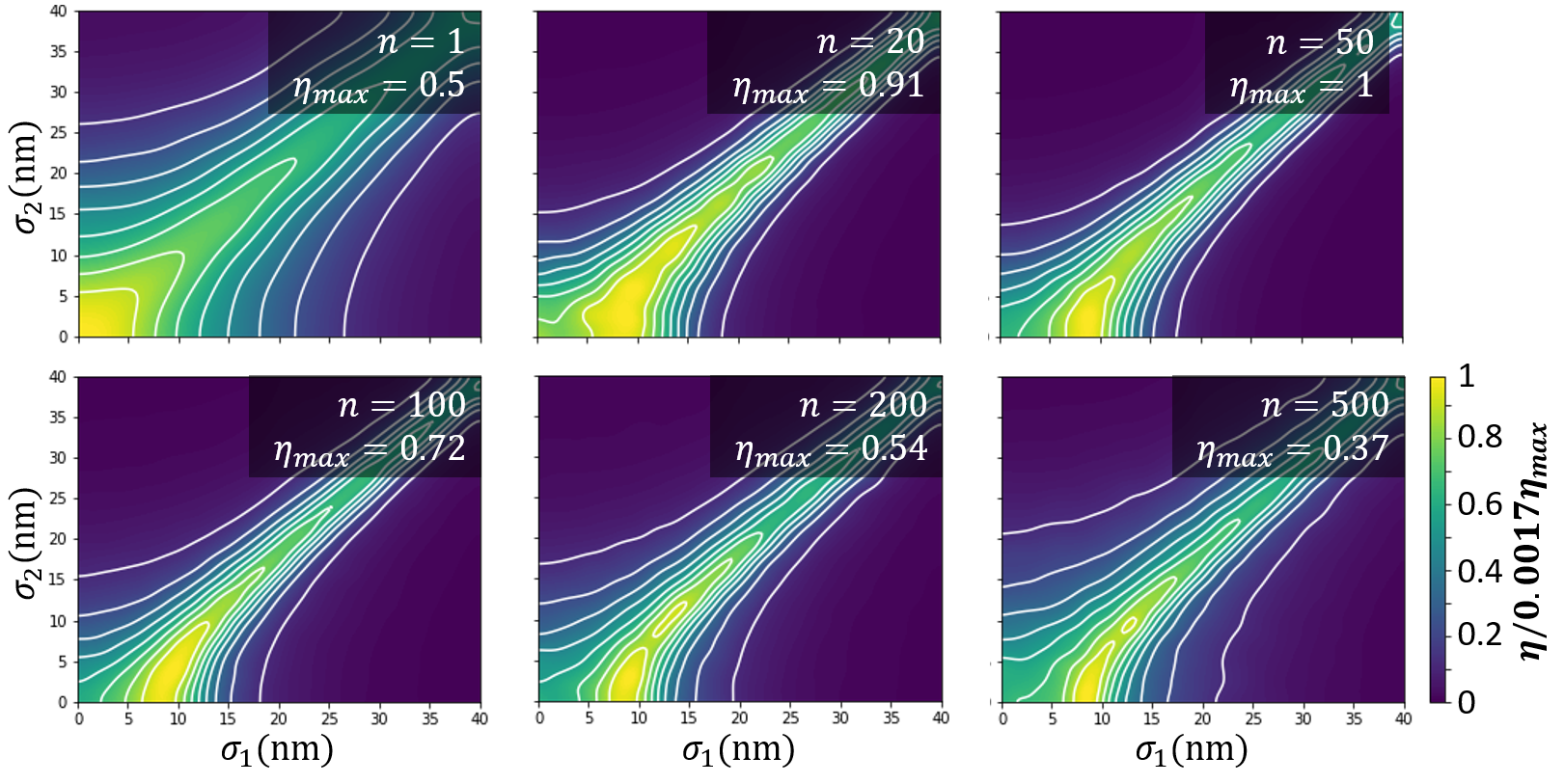}
    \caption{Normalized average asymmetry for a SiCu compound target with buried and rear surface gratings, and $\chi\left(\vec{q}\right)=\hat \chi \exp\left( - \sigma^2 q^2 - i\vec{r}_s \vec{q}\right)$, $\sigma^2=\sigma^2_2-\sigma_1^2$, with $\hat\chi=0.7$. We introduced $10\%$ normal distributed random variations in the ridge edge positions and heights, and varied the number $n$ of sub-volumina with different positions. }
    \label{fig:SiCuRepScan_noContaminants}
\end{figure*}

One prominent case certainly is that of a grating interface and grating rear surface, which we choose for a detailed discussion in the following. 
Such a target is both easy to produce and it enables a very easy analytical and numerical approach due to its periodicity. 
Here, we can assume noise being present in the form of random but small deviations from the perfect grating, i.e. each grating ridge is assigned to a random shift and width. 
We absorb the two into a random variation of the position of the left/right edge ridge $i$ by $\delta^i_{l/r}$ relative to the edges of the buried layer. 
In that case and if the edge position variations all are much smaller than the average grating ridge width $w$ itself, the replication factor can be written as
\begin{widetext}
\begin{equation}
    \chi_\delta\left(\vec{q}\right) = \chi_0\left(\vec{q}\right) \frac{\sum_k{\left[\sum_m{\left(e^{iq\left(w/2+\delta^{k,m}_{r,2}\right)}-e^{-iq\left(w/2-\delta^{k,m}_{l,2}\right)}\right)e^{i2kqw}}\right]}}{\sum_k{\left[\sum_m{\left(e^{iq\left(w/2+\delta^{k,m}_{r,1}\right)}-e^{-iq\left(w/2-\delta^{k,m}_{r,1}\right)}\right)e^{i2kqw}}\right]}}. 
    \label{eqn:randomShifts}
\end{equation}
\end{widetext}
which at the scattering peak positions $\vec{q}_N=N 2\pi/w$ and in small angle approximation simplifies to 
\begin{widetext}
\begin{equation}
    \chi_\delta\left(\vec{q}_N\right) \cong \chi_0\left(\vec{q}_N\right) \frac{\sum_k{\left[\sum_m{
    \frac{N\pi}{w}\left[(\delta_{l,2}^{k,m})^2-(\delta_{r,2}^{k,m})^2\right] + i (\delta_{r,2}^{k,m}+\delta_{l,2}^{k,m})}\right]}} {\sum_k{\left[\sum_m{
    \frac{n\pi}{w}\left[(\delta_{l,1}^{k,m})^2-(\delta_{r,1}^{k,m})^2\right] + i (\delta_{r,1}^{k,m}+\delta_{l,1}^{k,m})}\right]}} . 
\end{equation}
\end{widetext}
Here, the sum over $k$ goes over all grating ridges, the sum over m denotes the variation of parameters along the individual ridges. 
We first note that mere scale changes (i.e. $\delta^{k,m}_{r,i}=-\delta^{k,m}_{l,i}$) cancel out the real part of both the nominator and denominator, which means explicitly for real-valued $\chi_0$ that it remains real-valued and no asymmetry is generated. 
Additionally, as long as local shifts are symmetrically distributed, they cancel out, hence only a net shift of the center of mass averaged over all ridges can generate asymmetry, which then coincides with the former case of Eqn.~\eqref{eqn:shift} with $r_s$ substituted by the average net shift of the ridges. 
This can be seen in Fig.~\ref{fig:SiCuRepScan_noContaminants}, where we plot the average asymmetry as a function of the smoothness parameters $\sigma_1$ and $\sigma_2$ for different numbers $n$ of randomly distributed sub-volumina and $r_s=0$. 
Indeed, for $n\gg 1000$ the asymmetry vanishes. 
Then, the asymmetry is similar to that of a perfect grating. 
However, this is only true for very large values of $n$. 
For smaller $n$, the random shifts in fact do generate asymmetry, qualitatively and quantitatively similar to the case of a perfect grating but $r_s\ne 0$. 
\begin{figure*}
    \centering
    \includegraphics[width=\linewidth]{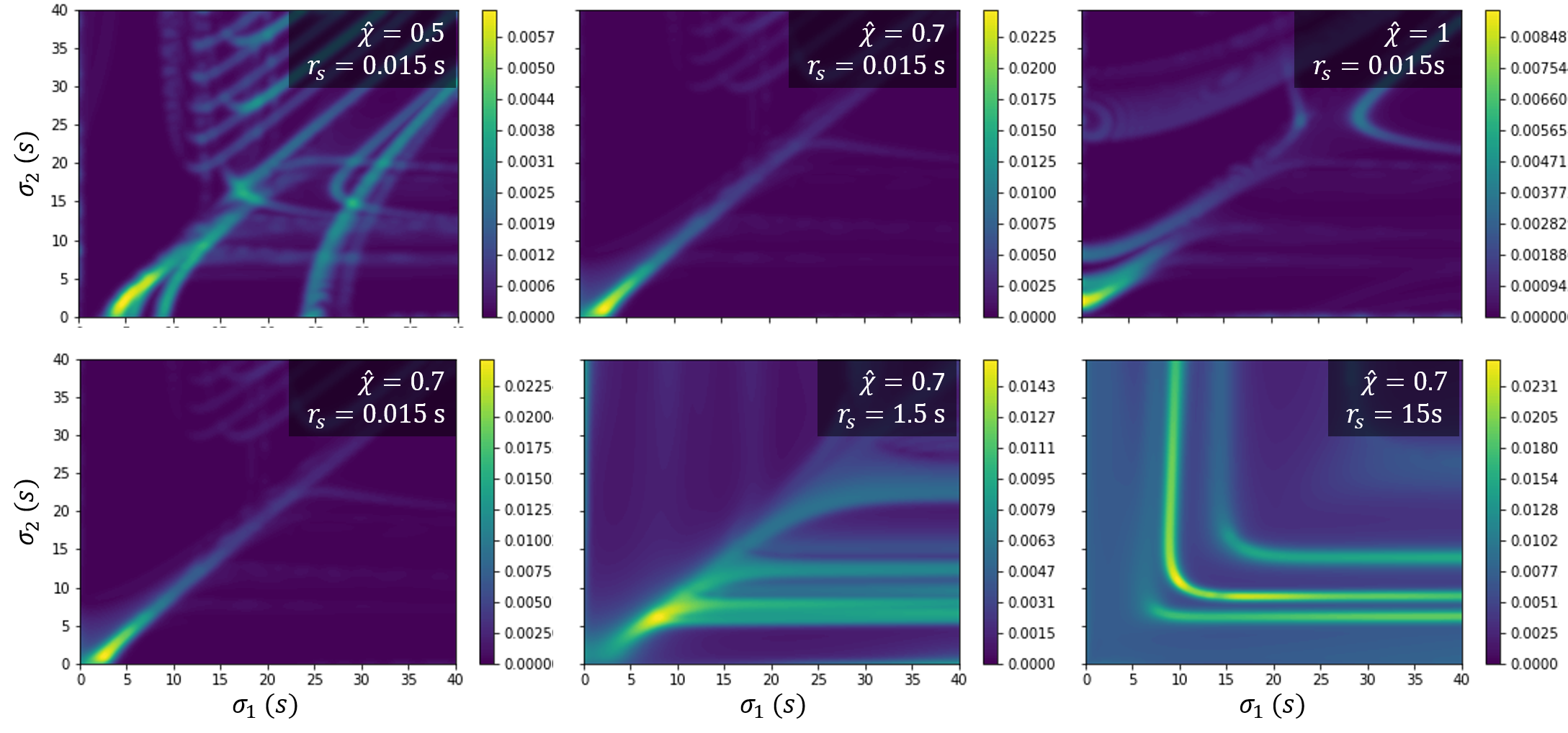}
    \caption{Average asymmetry for a SiCu compound target with buried and rear surface gratings, and $\chi\left(\vec{q}\right)=\hat \chi \exp\left( - \sigma^2 q^2 - i\vec{r}_s \vec{q}\right)$, $\sigma^2=\sigma^2_2-\sigma_1^2$, as before. We added CuO and hydro-carbon-oxide (thickness of both $5\%$ of $\hat z_0$) and SiO (thickness $0.3\%$ of $\hat z_0$) surface contamination layers to all applicable surfaces.\footnote{Those values corresponds to typical oxidation layer thicknesses ($\approx5\unit{nm}$ and $0.3\unit{nm}$ for Cu0 and SiO, respectively) and hydro-carbon-oxide layer thickness ($\approx5\unit{nm}$), when a grating height of $\hat{ z}_0=100\unit{nm}$ is adopted.} }
    \label{fig:SiCuSurface}
\end{figure*}
\\
\underline{Surface contamination and oxide layers}\\
A structurally different form of asymmetry is generated by another form of noise present in a realistic target: surface contaminations in the form of few nanometer thick carbohydroxyle contamination layers\cite{Belkind2008} or oxides on the surfaces of the materials forming a thin, yet inhomogeneous layer on the surfaces. 
Metals tend to oxidize quickly to oxidation layers thicknesses of a few nanometer, Cu e.g. with approximately $5\unit{nm}$\cite{SebastianMader2011}. 
Also silicon can oxidize to a thickness of a few atom layers\cite{NeergaardWaltenburg1995,*Hollauer2007}. 
Due to their small atomic number $Z^x$ they usually have a very small imaginary part of the optical correction at few $\keV$ X-ray photon energy. 
Having only negligible X-ray absorption in the multi-keV range, we limit ourself to the case of a contribution of the surface contamination layers to Thomson scattering only and ignore absorption, i.e. $f'=f''=0$, $f_0=Z^x$. 

We now want to add the effect that an oxidation layer or surface contamination may have on the asymmetry in Eqn.~\eqref{eqn:eta2layer}. 
As the contribution is again dependent on the geometry, we will again demonstrate this for the specific and practically relevant case of a grating surface.  
For simplicity we first neglect the structure factor and only consider a single grating ridge of width $w$
\begin{equation}
    z(x)=\frac{1}{2}\left(\mathrm{erf}\left(\frac{x+\frac{w}{2}}{\sqrt{2}\sigma}\right)-\mathrm{erf}\left(\frac{x-\frac{w}{2}}{\sqrt{2}\sigma}\right)\right).
\end{equation}
The oxidation layer projection $\Delta z^x(x)$ is then given, for a layer thickness $w^{x}$, approximately by
\begin{widetext}
    \begin{equation}
        \Delta z^x(x)=\xi\exp\left({-\frac{\left(x+\frac{1}{2}\left(w\pm w^{x}\right)\right)^2}{2\sigma^2}}\right)+\exp\left({-\frac{\left(x-\frac{1}{2}\left(w\pm w^{x}\right)\right)^2}{2\sigma^2}}\right)
    \end{equation}
\end{widetext}
with $\xi\cong 0.4 \left(\frac{w^{x}}{\sigma}\right)^{0.843}$. 
The plus sign refers to surface contaminations while the minus sign refers to oxidation layers. 
In Eqn.~\eqref{eqn:eta_start} we now have to add the term $\tilde{\zeta^x}=\rho^x\Delta\chi^x(q)$, where $\rho^x=Z^x n_i^x$ and $\Delta\chi^x(q)$ at the scattering peak positions reads
\begin{widetext}
$$ \Delta\chi^x(q) = \Delta \tilde{z}^x(q)/\tilde{z}_0\left(q\right) \cong \frac{ \xi\sigma q \cos\left(\frac{1}{2}q\left(w\pm w^{x}\right)\right)e^{-\frac{1}{2}q^2(\sigma^2-\sigma_0^2)}}{\sin{\left(\frac{qw}{2}\right)}}.$$ 
\end{widetext}
The random fluctuations can then be introduced in the same manner as carried out above in Eqn.~\eqref{eqn:randomShifts}. \\
Following the derivation of Eqn.~\eqref{eqn:eta2layer} above, we can give a similar equation for the asymmetry of a two-layer target including the surface layers, 
\begin{equation}
    \eta\left(\vec{q}\right)=\frac{\Im\left(c^x_0\Delta\chi^x\left(\vec{q}\right)+c^x_1    \chi\left(\vec{q}\right)\right)}{\Re\left[c^x_2 \left|\chi\left(\vec{q}\right)\right|^2+c^x_3 \chi\left(\vec{q}\right) + c^x_4\right]}
    \label{eqn:eta2layerNoise}
\end{equation}
where the constants $c^x_i$ are now not only material specific constants but rather depend also on the surface structure itself via $\Delta\chi^x(q)$ 
\begin{eqnarray*}
    c_0^x&=&2n_1n_xf''_1Z^x\\
    c_1^x&=&c_1\\
    c_2^x&=&c_2\\
    c_3^x&=&c_3+n_2 n_x (f^0_2+f'_2) Z^x \Delta\chi^x(q)\\
    c_4^x&=&c_4+2n_1n_x(f^0_1+f'_1)Z^x\Delta\chi^x(q)+(n_xZ^x)^2 \left|\Delta\chi^x(q)\right|^2. 
\end{eqnarray*}
The asymmetry from this equation is plotted in Fig.~\ref{fig:SiCuSurface} for different values of $\hat\chi$ and $r_s$ for a  Si-Cu sandwich target assuming a flat front side, a grating interface and a grating rear surface and no tilt, buried grating height $100\unit{s}$, grating period $300\unit{s}$. 
We find that especially $r_s$ has a large effect. 
While for small values of $r_s$ the pattern is similar to that without surface contaminations, albeit now much larger in absolute terms, for larger $r_s$ the pattern changes significantly. 

Again one can convince oneself that the imaginary part of $\Delta\chi^x$ is vanishing for $r_s=0$ and symmetric random fluctuations, so that again no asymmetry is expected if no tilt is present. 
As before, for values $n\lesssim 1000$ we find numerically that the asymmetry is not vanishing anymore. 
Rather, as we show below, a surprisingly stable asymmetry pattern can be found that depends only little on the random geometric variations but more pronounced on the geometry of the target layers, which therefore is accessible for measurements without the detailed knowledge of the noise and contamination layers. \\
The underlying reason for this effect is that the random variations of positions in fact do introduce an imaginary part of the replication factor - both of the target layers ($\chi$) as well as the surface contaminations ($\Delta\chi^x$). 
Its compensation depends on the illuminated target size and amount of randomness. 
If both are small, there simply is not enough variation in order to be fully compensated. 
We can conclude, that for very large illuminated target areas and small correlation lengths of the noise the asymmetry generated by the noise disappears and the asymmetry is given by Eqn.~\eqref{eqn:eta2layer}, \eqref{eqn:eta2layerSigma}, or \eqref{eqn:eta2layerNoise}. 

\begin{figure*}
    \centering
    \includegraphics[width=\linewidth]{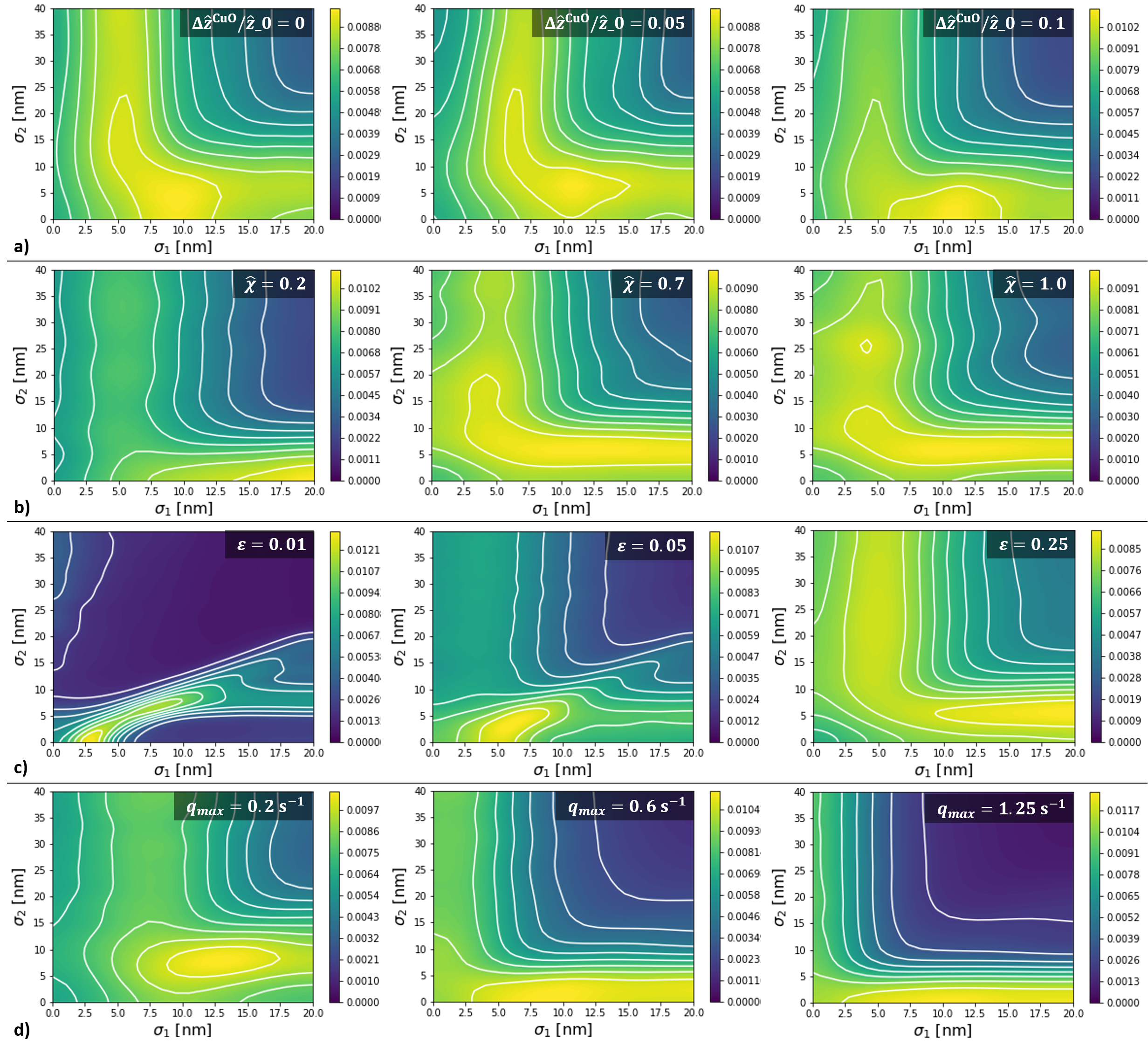}
    \caption{Average asymmetry in the $q$-range from $0-0.3 s^{-1}$ for a SiCu compound target with buried and rear surface gratings, and $\chi\left(\vec{q}\right)=\hat \chi \exp\left( - \sigma^2 q^2 - i\vec{r}_s \vec{q}\right)$, $\sigma^2=\sigma^2_2-\sigma_1^2$. We added CuO, hydro-carbon-oxide (both thickness $5\%$ of $\hat z_0$) and SiO (thickness $0.3\%$ of $\hat z_0$) surface contamination layers to all applicable surfaces, as before. We introduced $\varepsilon=10\%$ normal distributed random variations in the ridge edge positions, heights and layer thicknesses with $n=100$ sub-volumina with different positions. In a) we vary the thickness of the CuO oxidation layer; in b) we vary $\hat\chi$; in c) we vary $\varepsilon$; in d) we vary the q-range over which the average is taken. }
    \label{fig:SiCu__cC_rndm_Scan__Contaminants}
\end{figure*}
For all other cases the variation has to be taken into account numerically. 
In Fig.~\ref{fig:SiCu__cC_rndm_Scan__Contaminants} we show the numerically calculated asymmetry maps for the Si-Cu sandwich target. 
We vary the copper oxide layer thickness relative to the buried interface grating height, $\Delta\hat{z}^{CuO}/\hat{ z}_0$, the width $\varepsilon$ of the random gaussian distribution around the mean values of the grating ridge position offset (mean=0) and CuO layer thickness (mean=0.05), and the q-range over which the average of the asymmetry is taken. \\
Regardless of the specific values of the surface layers or the grating heights, we can find certain general laws:

§1: The asymmetry for a two material compound target with surface contaminations, flat front surface and given $\sigma_1$, $\sigma_2$ is a strong function of the randomness $\varepsilon$ and the $q$-range over which the asymmetry is averaged. 

§2: The choice of the `randomness' $\varepsilon$ has great impact on the asymmetry for small values of $\varepsilon$, while it converges to a stable pattern for larger values. 

§3: The asymmetry is largely independent on variations of the exact geometry, i.e. the pattern is stable with respect to different mean values of the relative surface layer thicknesses (see panel a) or ridge heights (b).

§4: The asymmetry with surface contaminations and small randomness is distributed qualitatively similar to the case without the surface contaminants, but much larger and blurred. 

§5: When $\varepsilon$ is increased to above $10\%$, the asymmetry pattern changes qualitatively similar to how it would change without randomness (Fig.~5), but now being blurred due to the randomness. 

§6: The asymmetry is largest if at least one of the gratings is sharp (but if the q-range is limited it shall not be ``too'' sharp). 
In that case, the asymmetry decreases if the sharp surface(s) expands/smoothes. 
This means for example that if initially one grating is smooth and the other one is sharp, a decrease of asymmetry can be used to measure the expansion of the latter, largely independent of the expansion of the former (if one assumes that the smoothness can only increase).